\documentclass[structabstract]{aa}  % ESTA ES LA OPCION PARA EL ORIGINAL
\usepackage{graphicx}
\usepackage{txfonts}
\begin{document}
   \title{Study of star-forming galaxies in SDSS up to redshift 0.4\\
II. Evolution from the fundamental parameters: mass, metallicity $\&$ SFR}

   \author{M.A. Lara-L\'opez
          \inst{1,2}
          \and
          A. Bongiovanni\inst{1,2}
	  \and
          J. Cepa\inst{1,2}
	  \and
	  A.M. P\'erez Garc\'{\i}a\inst{1,2}
	  \and
          M. S\'anchez-Portal\inst{3}
	  \and
          H. O. Casta\~neda\inst{1,3}
	  \and
          M. Fern\'andez Lorenzo\inst{1,2}
	  \and
          M. Povi\'c\inst{1,2}
         }

   \institute{Instituto de Astrof\'{\i}sica de Canarias, 38200 La Laguna, Spain
              \email{mall@iac.es}
         \and
             Departamento de Astrof\'{\i}sica, Universidad de la Laguna, Spain
          \and
            Departamento de F\'{\i}sica, Escuela Superior de F\'{\i}sica y Matem\'atica, IPN, M\'exico D.F., M\'exico
          \and
             Herschel Science Center, INSA/ESAC, Madrid, Spain
             }

   \date{Received; accepted}

\abstract  
  % context heading (optional)
{To understand the formation and evolution of galaxies, it is important to have a full comprehension of the role played by the metallicity, star formation rate (SFR), morphology, and color. The interplay of these parameters at different redshifts will substantially affect the evolution of galaxies and, as a consequence, the evolution of them will provide important clues and constraints on the galaxy evolution models. In this work we focus on the evolution of the SFR, metallicity of the gas, and morphology of galaxies  at low redshift in search of signs of evolution.} %leave it empty if necessary  
%   {}
 % aims heading (mandatory)
{To analyze the S2N2 (log({H$\alpha$}/[{S\,\textsc{ii}}]) vs. log({H$\alpha$}/[{N\,\textsc{ii}}])) diagram as a possible segregator of star--forming, composite, and AGN galaxies, to study the evolution of the Baldwin, Phillips $\&$ Terlevich (1981) diagrams, as well as the evolution of the SFR, metallicity, and morphology, through the mass--metallicity, luminosity--metallicity, SFR--stellar mass, and SFR--metallicity relationships of star--forming galaxies from SDSS--DR5 (Sloan Digital Sky Survey--Data Release 5), using redshift intervals in bins of 0.1 from $\sim$0 to 0.4.}
 % methods heading (mandatory)
{We used data processed with the STARLIGHT spectral synthesis code, correcting the fluxes for dust extinction, and estimating metallicities using the $R_{23}$ method. We use the S2N2 diagnostic diagram as a tool to classify star--forming, composite, and AGN galaxies. We analyzed the evolution of the three principal BPT diagrams, estimating the SFR and specific SFR (SSFR) for our samples of galaxies, studying the luminosity and mass-metallicity relations, and analyzing the morphology of our sample of galaxies through the $g-r$ color, concentration index, and SSFR.}
  % results heading (mandatory)
{We found that the S2N2 is a reliable diagram to classify star--forming, composite, and AGNs galaxies. We demonstrate that the three principal BPT diagrams show an evolution toward higher values of [{O\,\textsc{iii}}] $\lambda$5007/{H$\beta$} due to a metallicity decrement. We found an evolution in the mass--metallicity relation of $\sim$ 0.2 dex for the redshift range $0.3 < z < 0.4$ compared to our local one. From the analysis of the evolution of the SFR and SSFR as a function of the stellar mass and metallicity, we discovered a group of galaxies with higher SFR and SSFR at all redshift samples, whose morphology is consistent with those of late--type galaxies. Finally, the comparison of our local ($0.04<z<0.1$) with our higher redshift sample ($0.3<z<0.4$), show that the metallicity, the SFR and morphology, evolve toward lower values of metallicity, higher SFRs, and late--type morphologies for the redshift range $0.3<z<0.4$.}
  % conclusions heading (optional), leave it empty if necessary 
{}
% AA/2008/11033

\titlerunning{Evolution from fundamental parameters in SDSS galaxies}

\maketitle { }
%
%________________________________________________________________

\section{Introduction}

Baldwin, Phillips $\&$ Terlevich (1981, hereafter BPT) were the first to propose diagnostic diagrams to classify galaxies into starburst or active galactic nucleus (AGN), based on the dominant energy source in emission-line galaxies, since AGNs have a much harder ionizing spectrum than hot stars. Revised and refined by Veilleux $\&$ Osterbrock (1987), the three BPT empirical diagnostic diagrams use the optical line ratios [{O\,\textsc{i}}] $\lambda$6300/{H$\alpha$}, [{S\,\textsc{ii}}] $\lambda$$\lambda$6717, 6731/{H$\alpha$}, [{N\,\textsc{ii}}]  $\lambda$6583/{H$\alpha$}, and [{O\,\textsc{iii}}] $\lambda$5007/{H$\beta$} (hereafter  [{N\,\textsc{ii}}] will refer to [{N\,\textsc{ii}}] $\lambda$6583, and  [{S\,\textsc{ii}}] to [{S\,\textsc{ii}}] $\lambda$$\lambda$6717, 6731). The BPT diagrams are the most widely used method to segregate between star-forming galaxies and AGNs, since the lines in star--forming (SF) galaxies are emitted by {H\,\textsc{ii}} regions, which are ionized by massive stars, while AGNs are ionized by a harder radiation field.

Kewley et al. (2001, hereafter Kew01) used a combination of stellar population synthesis models and detailed self-consistent photoionization models to create a theoretical maximum starburst line on the three BPT diagrams. Kauffmann et al. (2003a, hereafter Kauf03) shifted this starburst limit to a lower and more precise one in the  [{N\,\textsc{ii}}]/{H$\alpha$} diagram, excluding Seyfert--{H\,\textsc{ii}} composite objects whose spectra contain significant contributions from both AGN and star formation, from pure SF galaxies. Galaxies between the Kauf03 and Kew01 divisions are considered as composite galaxies.

There are more division criteria between SF galaxies and AGNs in the [{N\,\textsc{ii}}]/{H$\alpha$} vs [{O\,\textsc{iii}}] $\lambda$5007/{H$\beta$} BPT diagram, such as the one of Stasi\'nska et al. (2006), which used a limit lower than that of Kauf03, based on a more rigurous criterion, and the one of Lee et al. (2007), which used an intermediate empirical  line between the Kauf03 and  Kew01 divisions. It is possible, however, to classify SF galaxies and AGNs  using only the [{N\,\textsc{ii}}]/{H$\alpha$} ratio, as discussed in Stasi\'nska et al. (2006), since the left arm of the [{N\,\textsc{ii}}]/{H$\alpha$} diagram (see for example Fig. 2) is a measure of the combination of the metallicity and the ionization parameter. Then, larger values of this ratio indicate that the galaxy host an AGN. Stasi\'nska et al. (2006) classify as SF galaxies those with log([{N\,\textsc{ii}}]/{H$\alpha$}) $\leq$ -0.4, composite galaxies those with -0.4 $<$ log([{N\,\textsc{ii}}]/{H$\alpha$}) $\leq$ -0.2, and AGNs those galaxies with log([{N\,\textsc{ii}}]/{H$\alpha$}) $>$ -0.2.

Following with the objective of segregate SF from composite and AGNs galaxies, in this work we study the S2N2 diagram as a reliable segregator of galaxies. This log({H$\alpha$}/[{S\,\textsc{ii}}]) vs. log({H$\alpha$}/[{N\,\textsc{ii}}])  diagram was introduced by Sabbadin et al. (1977) as a useful tool  to separate galactic planetary nebula (PNe), {H\,\textsc{ii}} regions, and supernova remnants (SNRs). This diagram was later applied to Herbig-Haro objects (Cant\'o 1981), Galactic PNe (Garc\'{\i}a-Lario et al. 1991, Riesgo $\&$ L\'opez 2005), and extragalactic PNe (Magrini et al. 2003). The S2N2 diagram has been used also as a metallicity and ionization parameter indicator for extragalactic {H\,\textsc{ii}} regions by Viironen et al. (2007).

The S2N2 diagram has been also applied to galaxies by some authors. For example, Moustakas  $\&$  Kennicutt (2006) studied whether there was a difference between integrated spectra of galaxies and the spectra of individual {H\,\textsc{ii}} regions. Dopita et al. (2006)  used the S2N2 diagram, among others, for abundance diagnostics using photoionization models. Nevertheless, the [{S\,\textsc{ii}}] flux shows always deficiences when generated by photoionization models (e.g. Levesque et al. 2010). Also, Lamareille et al. (2009) and P\'erez-Montero et al. (2009) used the S2N2 diagram as a segregator of SF from Seyfert 2 galaxies, but using different ratios: log([{N\,\textsc{ii}}]/{H$\alpha$}) vs log([{S\,\textsc{ii}}]/{H$\alpha$}). However, in their division Lamareille et al. (2009) do not distinguish between SF and composite galaxies, also, they used equivalent widths instead of emission line fluxes, which could affect the results (Kobulnicky $\&$ Kewley 2004).

The formation and evolution of galaxies at different cosmological epochs are driven mainly by two linked processes: the star formation history and the metal enrichment. Thus, from an observational point of view, the star formation rate (SFR), the metallicity and the stellar mass of the galaxies at different epochs will give us important clues on the evolution of galaxies. The first quantitative SFRs were derived from evolutionary synthesis models of galaxy colors (Tinsley 1968, 1972, Searle et al. 1973), confirming the trends in SFRs and star formation histories along the Hubble sequence, and giving the first predictions of the evolution of the SFR with cosmic lookback time. The development of more precise direct SFR diagnostics includes the integrated emission--line fluxes (Cohen 1976, Kennicutt 1983), near-ultraviolet continuum fluxes (Donas $\&$ Deharveng 1984), and infrared continuum fluxes  (Harper $\&$ Low 1973, Rieke $\&$ Lebofsky 1978, Telesco $\&$ Harper 1980); see Kennicutt (1998) for a review. The hydrogen Balmer line {H$\alpha$}  is currently the most reliable tracer of star formation, since  in {H\,\textsc{ii}} regions and star-forming galaxies, the Balmer emission-line luminosity scales directly with the total ionizing flux of the embedded stars. A widely known calibration of the {H$\alpha$} line as SFR tracer is the one devised by Kennicutt (1998). However, it is important to take into account corrections for stellar absorption and reddening to obtain SFRs in agreement with the ones derived using other wavelengths (e.g. Rosa-Gonz\'alez et al. 2002, Charlot et al. 2002, Dopita et al. 2002). In parallel, other diagnostics have been developed using the oxygen doublet [{O\,\textsc{ii}}] $\lambda$3726, 3729 for the redshift range $z \sim 0.4-1.5$ (e.g. Gallagher et al. 1989, Kennicutt 1998, Rosa-Gonz\'alez et al. 2002, Kewley et al. 2004). Moreover, this diagnostic is usefull when the {H$\alpha$} line is not easily observable at higher redshifts ($z \gtrsim 0.4$ in the optical). However, the [{O\,\textsc{ii}}] doublet presents problems in reddening and abundance dependence (Jansen, Franx $\&$ Fabricant 2001, Charlot et al. 2002). Alternatively, it is possible to estimate the SFR from the soft X-ray luminosity, which is comparable to that determined from the {H$\alpha$} luminosity (Rosa Gonz\'alez et al. 2009, Rovilos et al. 2009).

A strong dependence of  the SFR and the stellar mass and its evolution with redshift has been found, with the bulk of star formation occurring first in massive galaxies, and later in less massive systems (e.g. Guzm\'an et al. 1997, Brinchmann $\&$ Ellis 2000, Juneau et al. 2005, Bauer et al. 2005, Bell et al. 2005, P\'erez-Gonzalez et al. 2005, Feulner et al. 2005, Papovich et al. 2006, Caputi et al. 2006, Reddy et al 2006, Erb et al. 2006, Noeske et al. 2007a, Buat et al. 2008). In the local universe, several studies have illustrated a relationship between the SFR and stellar mass, identifying two populations: galaxies on a star-forming sequence, and $``$quenched" galaxies, with little or no detectable star formation (Brinchmann et al. 2004, Salim et al. 2005, Lee 2006). At higher redshift, Noeske et al. (2007a) showed the existence of a $``$main sequence" (MS) for SF galaxies in the SFR--stellar mass relation over the redshift range $0.2 < z < 1.1$. From the galaxies considered in this study, it was shown that the slope of the MS remains constant to $z>1$, while the MS as a whole moves to higher SFR as $z$ increases.

Metallicity is another important property of galaxies, and its study is crucial for a deep understanding of  galaxy formation and evolution, since it is related to the whole past history of the galaxy. Metallicity is a tracer of the fraction of baryonic mass already converted into stars and is sensitive to the metal losses due to stellar winds, supernovae and active nuclei feedbacks. A detailed description of the different metallicity methods and calibrations are given in Lara-L\'opez et al. (2009a,b).

Stellar mass and metallicity are strongly correlated in SF galaxies, with massive galaxies showing higher metallicities than less massive galaxies. This relationship provides crucial insight into galaxy formation and evolution. The mass-metallicity ($M-Z$) relation was first observed by Lequeux et al. (1979), has been intensively studied (Skillman et al. 1989; Brodie $\&$ Huchra 1991; Zaritsky et al. 1994; Richer $\&$ McCall 1995; Garnett et al. 1997; Pilyugin $\&$ Ferrini 2000, among others), and it is well established by the work of Tremonti et al. (2004, hereafter T04) for the local universe (z $\sim$ 0.1) using SDSS data. The study of the redshift evolution of the $M-Z$ relation has provided us with crucial information on the cosmic evolution of star formation.

Regarding the evolution of the $M-Z$ relation for SF galaxies at z $<$ 1, Savaglio et al. (2005), have investigated the mass--metallicity relations using galaxies at 0.4 $<$ z $<$ 1, finding that metallicity is lower at higher redshift by $\sim$ 0.15 dex. Moreover, Maier at al. (2005), Hammer et al. (2005), and Liang et al. (2006) found that emission line galaxies were poorer in metals at z $\sim$ 0.7 than present--day spirals. A study of Lamareille et al. (2009) focused on the evolution of the $M-Z$ relation up to z $\sim$ 0.9, suggesting that the $M-Z$ relation is flatter at higher redshifts. However, Carollo $\&$ Lilly (2001), from emission--line ratios of 15 galaxies in a range of 0.5 $<$ z $<$ 1, found that their metallicities appear to be remarkably similar to those of local galaxies selected with the same criteria. Also, Lilly et al. (2003), from a sample of 66 SF galaxies with 0.47 $<$ z $<$ 0.92, claim a smaller variation in metallicity of $\sim$ 0.08 dex compared with the metallicity observed locally, showing only modest evolutionary effects (for more details about the $M-Z$ relation, see Lara-L\'opez et al. 2009b).

In a recent study, Calura et al. (2009) have demonstrated the importance on the morphology of galaxies when deriving the $M-Z$ relation since, at any redshift, elliptical galaxies present the highest stellar masses and the highest metallicities, whereas the irregulars are the least massive galaxies, characterised by the lowest O abundances.

In this paper, we consistently approach several topics, starting with the introduction of the S2N2 as a reliable diagram to classify galaxies, the analysis of the metallicity evolution of galaxies in the three BPT diagrams, and, for a better understanding of the processes involved in the observed evolution of galaxies at low redshift, we studied the mass, metallicity and SFR relations, such as the $M-Z$, metallicity-SFR and mass-SFR relations. We also point out that the morphology of galaxies play an important role when deriving conclusions, since late--type galaxies will result in lower metallicity estimates and higher SFRs than early--type (Calura et al. 2009).

This paper is structured as follows, in Sect. 2 we detail the data used for this study, the dust extinction correction and the metallicity estimates for our sample of galaxies, in Sect. 3 we introduce the S2N2 as a reliable diagram to segregate SF, composite, and AGNs galaxies. In Sect. 4  we analyzed the evolution of the BPT diagrams. In Sect. 5 we investigate the evolution of the mass-metallicity and luminosity-metallicity relations, whereas in Sect. 6 we discuss the relations between the SFR and SSFR with stellar mass and metallicity, as well as the morphology of our galaxies using colors, concentration index, and SSFRs. Finally, conclusions are given in Sect. 7.

%__________________________________________________________________

\section{Data processing and sample selection}
%                                     Two column figure (place early!)
%______________________________________________ Gamma_1 (lg rho, lg e)

We selected emission line galaxies from SDSS--DR5 (Adelman--McCarthy et al. 2007). Data were taken with a 2.5 m telescope located at Apache Point Observatory (Gunn et al. 2006). The SDSS spectra were obtained using 3 arcsec diameter fibres, covering a wavelength range of 3800-9200 {\AA}, and with a mean spectral resolution $\lambda$/$\Delta\lambda$ $\sim$ 1800. The SDSS--DR5 spectroscopy database contains spectra for $\sim$ $\rm{10}^6$ objects over $\sim$ 5700 $\rm{deg}^2$. Further technical details can be found in Stoughton et al. (2002).

We used the SDSS--DR5 spectra from the STARLIGHT database\footnote{http://www.starlight.ufsc.br}, which were processed with the STARLIGHT spectral synthesis code, developed by Cid Fernandes and collaborators (Cid Fernandes et al. 2005, 2007, Mateus et al. 2006, Asari et al. 2007). From the spectra, the STARLIGHT code subtracts the continuum, obtaining the emission lines fluxes measurements for each galaxy. For each emission line, the STARLIGHT code returns the rest frame flux and its associated equivalent width, linewidth, velocity displacement relative to the rest--frame wavelength and the S/N of the fit. In the case of Balmer lines, the underlying stellar absorption was corrected by the STARLIGHT code using synthetic spectra obtained by fitting an observed spectrum  with a combination of 150  simple stellar populations (SSPs) from the evolutionary synthesis models of Bruzual $\&$ Charlot (2003).

From the full set of galaxies, we only consider galaxies whose spectra show in emission the  {H$\alpha$}, {H$\beta$}, [{N\,\textsc{ii}}], [{O\,\textsc{ii}}] $\lambda$3727, [{O\,\textsc{iii}}] $\lambda$4959, [{O\,\textsc{iii}}] $\lambda$5007, [{O\,\textsc{i}}] $\lambda$6300, [{S\,\textsc{ii}}]  lines. We selected galaxies with a signal--to--noise ratio higher than 3$\sigma$ for the {H$\alpha$}, {H$\beta$}, and [{N\,\textsc{ii}}] lines.

In order to identify any  evolution of galaxy parameters or relations, we divided our sample in four redshift intervals as follow: $0.04 \leq z_0 <  0.1$, $0.1 \leq z_1 <  0.2$, $0.2 \leq z_2 <  0.3$, $0.3 \leq z_3 \leq  0.4$. The lower limit of $z_0$ corresponds to an aperture covering fraction of 20$\%$, which is the minimum required to avoid domination of the spectrum by aperture effects (Kewley et al. 2005). This classification give us 85931 galaxies for $z_0$, 48888 galaxies for $z_1$, 3278 galaxies for $z_2$, and 199 galaxies for $z_3$.

We selected galaxies with an apparent Petrosian $r$ magnitude of   $14.5 < r < 17.77$ in the redshift samples $z_0$, $z_1$, and $z_2$, which yields 82884, 44763, and 1802 galaxies, respectively, corresponding to the magnitude completeness at these redshifts (see Fig. 1). Galaxies of the $z_3$ sample have a different completeness range 16.9 $< r <$ 18.8, as observed in Fig. 1, giving 119 galaxies. We used the $z_0$ and $z_1$ sample of galaxies with its respective completeness, but for galaxies of samples $z_2$ and $z_3$ we used both, those in the completeness range, and those out of the completeness range. The reason for this, is to improve the galaxy statistics by increasing the number of them. As we shown in the next sections, the main results are similar using galaxies in the magnitude completeness and galaxies of the total sample.

\begin{figure}[ht!]
\centering
\includegraphics[scale=0.80]{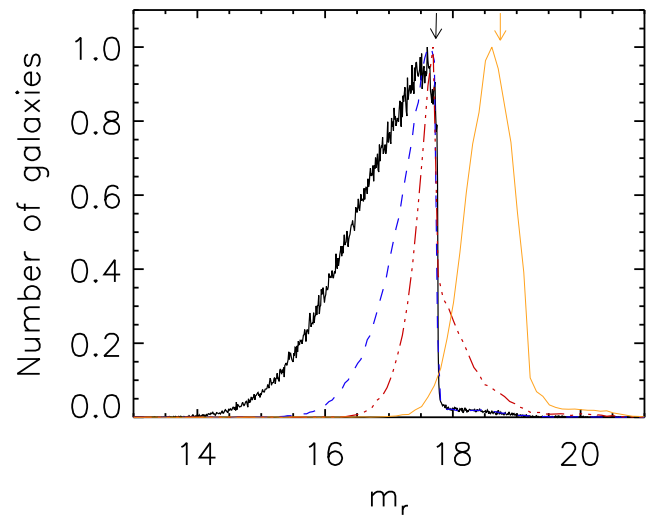}
\caption{Normalized histogram of the apparent Petrosian $r$ magnitudes in the four redshift bins. Dark solid line represents galaxies at $z_0$, dashed line the galaxies at $z_1$, dot-dashed lines the galaxies at $z_2$, and clear solid line, the galaxies at $z_3$. The black arrow shows the completeness limit for the samples $z_0$, $z_1$, and $z_2$, and clear arrow points the same for $z_3$.}
\end{figure}

\subsection{Sample selection for Section 3}

In Section 3, to study the S2N2 diagram as a segregator of different types of galaxies, we adopt the  $``$main galaxy sample" (e.g. Strauss et al. 2002) with Petrosian $r$ magnitudes in the range 14.5 $< r <$ 17.77, and the redshift interval $z_0$, taking into account all the emission lines and the signal--to--noise ratio mentioned above, which yields 82884 galaxies. In this Section, we used AGN, composite, and SF galaxies.

\subsection{Sample selection for Section 4}

In Section 4, we study the evolution of the three BPT diagrams using the four redshift intervals, magnitude intervals, and the signal--to--noise restrictions mentioned above. In this Section, we used AGN, composite, and SF galaxies. Also, to study the metallicity evolution of the SF galaxies of the BPT diagrams (see Fig. 6), we used the sample of Sections 5 and 6, mentioned below.

\subsection{Sample selection for Sections 5 and 6}

For Sections 5 and 6, we selected SF galaxies following the criterion given by Kauf03 in the BPT empirical diagnostic diagram:  log[{O\,\textsc{iii}}] $\lambda$5007/{H$\beta$} $\le$ 0.61/$\{$log([{N\,\textsc{ii}}] /{H$\alpha$})-0.05$\}$ + 1.3, the same used by Veilleux $\&$ Osterbrock (1987), Kewley et al. (2001, 2006), and Stasi\'nska et al. (2006), among others. After all these selections, the number of galaxies of each redshift bin is reduced to 61921 SF galaxies for $z_0$, 27853 for $z_1$, 1671 for $z_2$, and 67 {H\,\textsc{ii}} galaxies for $z_3$.

The extinction correction and metallicity estimates were calculated as in Lara-L\'opez et al. (2009b). Our extinction correction was derived using the Balmer decrements in order to obtain the reddening coefficient C({H$\beta$}). We used the Cardelli et al (1989) law, with $R_v$ = $A_v$/E(B -- V)=3.1, assuming case B recombination with a density of 100 cm$^{-3}$ and a temperature of 10$^{4}$ K, with {H$\alpha$}/{H$\beta$} $=$ 2.86 (Osterbrock 1989).

We estimated metallicities using the $R_{23}$ relation introduced by Pagel et al. (1979), $R_{23}$=([{O\,\textsc{ii}}] $\lambda$3727+[{O\,\textsc{iii}}] $\lambda\lambda$4959, 5007)/H$\beta$, and adopted the calibration given by Tremonti et al. (2004), $12+\rm{log(O/H)} = 9.185 - 0.313x - 0.264x^2 - 0.321x^3$, where $x$ = log $R_{23}$. We selected the upper branch  of the double--valued $R_{23}$, in which the Tremonti et al. (2004) calibration is valid, taking 12$+$log(O/H) $>$ 8.4 and log([{N\,\textsc{ii}}]/[{O\,\textsc{ii}}]$\lambda$3727) $>$ --1.2, since the upper and lower branches of the $R_{23}$ calibration bifurcates at those values (see Kewley $\&$ Ellison 2008).

Applying this final criterion, we end with 58866 galaxies for $z_0$, 24385 for $z_1$, 1631 for $z_2$ (from which 712 galaxies are in their completeness magnitude interval), and 62 galaxies for $z_3$ (from which 41 galaxies are in their completeness magnitude interval), all of them in the upper branch of the $R_{23}$ relation, corresponding to the $\sim$99 $\%$ to the {H\,\textsc{ii}} classified galaxies. Then, we are not introducing a bias selecting the upper branch for this samples.

\section{The S2N2 diagnostic diagram as a star--forming, composite, and  AGN galaxies segregator} 

As mentioned in Section 1, BPT diagrams are the most used method to segregate between star-forming and AGN galaxies.  From the three BPT diagrams ([{N\,\textsc{ii}}]/{H$\alpha$}, [{S\,\textsc{ii}}]/{H$\alpha$}, and [{O\,\textsc{i}}] $\lambda$6300/{H$\alpha$} vs [{O\,\textsc{iii}}] $\lambda$5007/{H$\beta$}), the most used is the [{N\,\textsc{ii}}]/{H$\alpha$} vs [{O\,\textsc{iii}}] $\lambda$5007/{H$\beta$} one, since it is the only one that can segregate pure SF and composite galaxies, as demonstrated by Kewley et al. (2006) and P\'erez-Montero et al. (2009). The  other two BPT diagrams are not useful for segregating SF from composite objects.

\begin{figure}[ht!]
\centering
\includegraphics[scale=0.65]{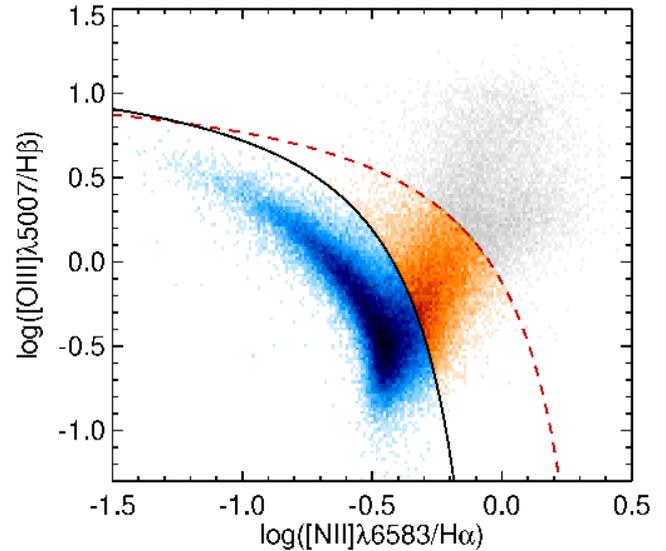}
\caption{Log([{N\,\textsc{ii}}]/{H$\alpha$}) vs log([{O\,\textsc{iii}}] $\lambda$5007/{H$\beta$}) BPT diagram. Solid line show the Kauf03 empirical division between SF and composite galaxies, and dashed line represents the Kew01 starburst limit. [See the electronic edition of the Journal for a color version of this figure.]}
\end{figure}

Commonly known as the  S2N2, the log({H$\alpha$}/[{S\,\textsc{ii}}]) vs. log({H$\alpha$}/[{N\,\textsc{ii}}])  diagram  has been used to separate planetary nebulae (PNe), {H\,\textsc{ii}} regions, and supernova remnants (SNRs, see Sabbadin et al 1977, Riesgo $\&$ L\'opez 2006, Viironen et al. 2007). We propose the S2N2 diagram to classify SF, composite, and AGN galaxies, something until now only possible  with the  [{N\,\textsc{ii}}]/{H$\alpha$} vs [{O\,\textsc{iii}}] $\lambda$5007/{H$\beta$} diagram. However, the S2N2 diagram use only the {H$\alpha$}, [{N\,\textsc{ii}}], and [{S\,\textsc{ii}}] emission lines, all of them close in wavelength, avoiding reddening corrections, and making possible its use for surveys limited in spectral range.

From the  $``$main galaxy sample", we consider SF galaxies as those lying below the Kauf03 division, composite galaxies as those lying between the Kauf03 and Kew01 lines, and AGNs galaxies, those above the Kew01 division (see Fig. 2). From the total sample (82884 galaxies, see Sect. 2), the 71.4$\%$, 19$\%$, and 9.6$\%$ correspond to SF, composite, and AGN galaxies, respectively. Taking this classification as a reference, we plotted the S2N2 diagram with the three classifications of galaxies (see Fig 3 and 4).

\begin{figure}[ht!]
\centering
\includegraphics[scale=0.40]{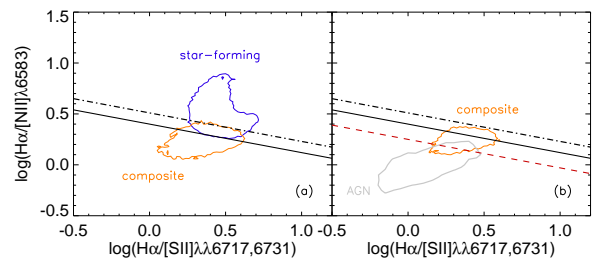}
\caption{Contour plots for star--forming, composite, and AGN galaxies. ($a$) Star--forming, darker (blue) contour, and composite, lighter (orange) contour, enclosing the $\sim$90$\%$ ($\sim$1.8 $\sigma$) of each sample. The solid line establish our limit for star--forming galaxies, while dot--dashed line delimit the almost pure star--forming galaxies. ($b$) Composite, darker (orange) contour, and AGN, lighter (grey) contour, enclosing $\sim$75$\%$ ($\sim$1.2 $\sigma$) of each sample. The dashed line shows our separation for composite and AGN galaxies, solid and dot--dashed lines as in panel ($a$). [See the electronic edition of the Journal for a color version of this figure.]}
\end{figure}

\begin{figure*}[ht!]
\centering
\includegraphics[scale=0.65]{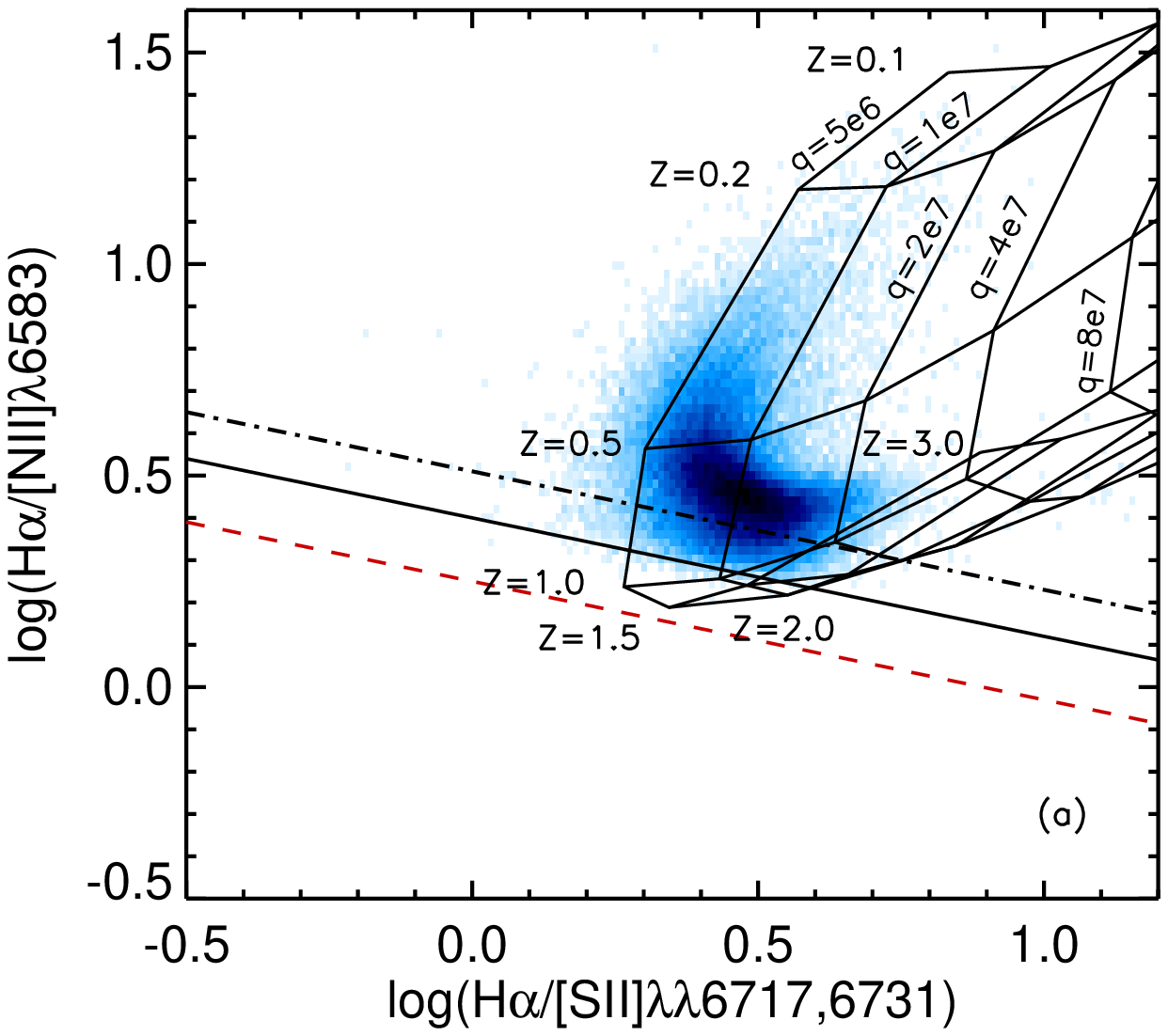}
\includegraphics[scale=0.65]{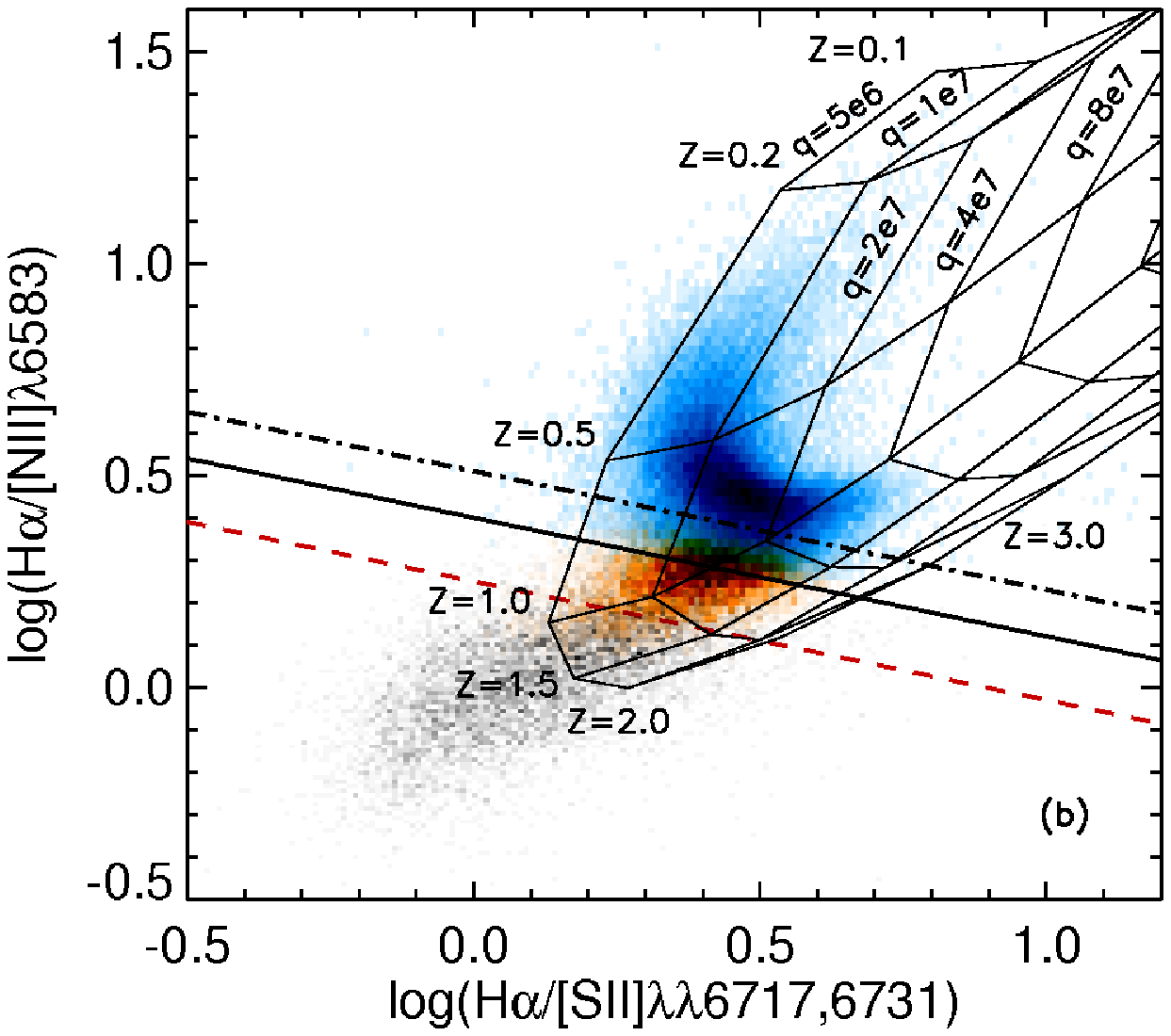}
\caption{($a$) log({H$\alpha$}/[{S\,\textsc{ii}}] ) vs. log({H$\alpha$}/[{N\,\textsc{ii}}]) diagram, corresponding to the star--forming galaxies below the Kauf03 separation in the BPT diagram. We overploted the grid of the photoionization models of Kewley et al. (2001) created with PEGASE, taking an instantaneous burst model with an electronic density of 10 cm$^{-3}$. Dot-dashed, solid, and dashed line, represent our division for pure star--forming and composite, star--forming and composite, and composite and AGN galaxies, respectively. ($b$) Same diagram with the grid of the photoionization models of Kewley et al. (2001) created with PEGASE, taking a continuous model with an electronic density of 10 cm$^{-3}$. [See the electronic edition of the Journal for a color version of this figure.]}
\end{figure*}

In order to establish division lines to separate SF, composite and AGN galaxies in the S2N2 diagram, we generated contour plots for each category of galaxies (see Fig. 3). As our sample of galaxies is larger for the SF and composite galaxies, we used contours enclosing $\sim$90$\%$ for those galaxies. However, as AGN galaxies are less numerous, we used contours enclosing $\sim$75$\%$ for composite and AGNs. The contour plots shown in  Fig. 3a delimit two tangent parallel lines, generating with this criterium division lines defined by Eqs. (1) and (2). In order to define a division line between composite and AGN galaxies,  we sampled the plot area with parallel lines of Eq. (1) in bins of 0.02 dex, generating in this way histograms for composite and AGN galaxies, where Eq. 3 corresponds to the intersection of both histograms.\\

 (i)  \emph{Pure star--forming galaxies} are separated by:

\begin{equation}
{\rm log}({\rm{H\alpha}}/[${N\,\textsc{ii}}$]) > -0.28 \;{\rm x}\;{\rm log({H\alpha}}/[${S\,\textsc{ii}}$])+0.51, 
\end{equation}  98.8$\%$ of galaxies above this line are SF galaxies and correspond to  88$\%$ of the SF sample. \\

(ii)  \emph{Star--forming and composite galaxies} are divided by:

\begin{equation}
{\rm log}({\rm{H\alpha}}/[${N\,\textsc{ii}}$]) > -0.28 \;{\rm x}\; {\rm log({H\alpha}}/[${S\,\textsc{ii}}$])+0.40, 
\end{equation}  99.8$\%$ of the SF sample lay above this line. However, from all galaxies above this line, 8$\%$ will correspond to composite galaxies, and  the remaining 91$\%$ to SF galaxies. \\

(iii) \emph{Composite and AGN galaxies} are divided by:

\begin{equation}
{\rm log}({\rm{H\alpha}}/[${N\,\textsc{ii}}$]) > -0.28 \;{\rm x}\; {\rm log({H\alpha}}/[${S\,\textsc{ii}}$])+0.25, 
\end{equation} 74.6$\%$ of the galaxies below this line are AGNs. Unfortunately, this diagram does not allow separating Seyfert from Liner galaxies (see Fig. 3). Composite galaxies are selected as those between the lines of Eq. (1) and (3). From all galaxies between both lines, 61.2$\%$, 33.2$\%$, and 5.6$\%$ correspond to composite, SF, and AGN galaxies, respectively.

In table 1 we compare the N2 ratio, considering the Stasi\'nska limits, with the S2N2 diagram in order to test their respective ability for segregating SF, composite, and AGN galaxies. This table shows the percentage of galaxies classified according to each division line, as well as its corresponding contamination and percentage of missed galaxies taking as a reference the classification of the log([{N\,\textsc{ii}}]/{H$\alpha$}) vs log([{O\,\textsc{iii}}] $\lambda$5007/{H$\beta$}) diagram. As can be appreciated for the SF segregation, the N2 division miss the highest percentage of SF galaxies, while our pure star--forming division (Eq. 1), although with the same contamination from composite galaxies, miss less SF galaxies. Moreover, our SF limit (Eq. 2) enclose almost all SF galaxies with a small contamination from composite galaxies. This provides the additional advantage with respect to the N2 diagram, that the user can choose between the possibilities of selecting either most SF galaxies, or galaxies with the smallest contamination from composite galaxies. The composite division for the N2 ratio shows a similar percentage of composite and SF galaxies. However, the S2N2 division, allows obtaining a higher percentage of composite galaxies, missing less composite galaxies, with a lower contamination of SF galaxies, and with only a quite small increment of AGN contamination. For segregating AGNs, while the S2N2 diagram provides a lower contamination from composite galaxies, miss $\sim$6$\%$ more AGNs than using the N2 ratio.

\begin{table*}[t]
\begin{center}
\begin{tabular}{c|ccc|cccc|ccc}   
\hline
\hline
\multicolumn{1}{c} { }&\multicolumn{3}{c} {SF segregation ($\%$)}&\multicolumn{4}{c} {Composite segregation ($\%$)}&\multicolumn{3}{c}{AGN segregation ($\%$)}\\\cline{2-11}
\hline
 &SF &SF-missed &Comp&Comp&Comp-missed&SF&AGN&AGN&AGN-missed&Comp \\\hline
N2&98.8&18.6&1.2&49.3&27.3&47.7&3&66&9&34\\\
S2N2(1$\&$3)&98.8&11.8&1.2&61.2&18.6&33.2&5.6&74.6&15.3&25.4\\\
S2N2(2)&91&0.2&8&&&&&&&\\\hline
\noalign{\smallskip}
\end{tabular}
\normalsize
\rm
\end{center}
\caption{Comparison between the S2N2 diagram and the N2 ratio to segregate SF, composite, and AGN galaxies. S2N2(1$\&$3) and S2N2(2) refer to galaxies taking as a reference Eq. (1$\&$3) and Eq. (2), respectively. The SF block indicate for the S2N2(2) division, for example, that taking all galaxies above Eq. (2), the 91$\%$ will correspond to SF galaxies, and the rest 8$\%$ to composite galaxies, also, that we have missed 0.2$\%$ of the original SF sample. In the case of the composite block, for the S2N2(1$\&$3) division we are taking galaxies between the lines defined by Eqs. (1) and (3).}
\end{table*}

In Fig. 4, we overplot the pre--run photoionization grids of Dopita et al. (2000) and Kew01 for an instantaneous burst model and for a continuous starburst model. The best grid to our SF galaxies for the S2N2 diagram is the corresponding to an instantaneous burst model, with an electronic density of 10 cm$^{-3}$, using the PEGASE code (see Fig. 4a). As explained in Dopita et al. (2000), the high surface brightness isolated extragalactic {H\,\textsc{ii}} regions are in general excited by young clusters of OB stars and that, in this case, the ionizing EUV spectra and {H\,\textsc{ii}} region emission-line spectra predicted by the PEGASE and STARBURST99 codes for an instantaneous, zero-age star formation model, are essentially identical.

In their work, Dopita et al. (2000) and Kew01 modeled a large sample of infrared starburst galaxies using both the PEGASE v2.0 (Fioc $\&$ Rocca-Volmerange 1977) and the STARBURST99 (Leitherer et al. 1999) codes to generate the spectral energy distribution (SED). In both cases, MAPPINGS III code was used to compute photoionization models. The pre-run grids use photoionization models with  ionization parameters $q$ (cm s$^{-1}$) in the range $5 {\rm x}10^{6} \leq q \leq 3 {\rm x}10^{8}$, and metallicities from Z=0.05 to 3 Z$_{\odot}$; moreover, two values for electronic density were used, 10 and 350 cm$^{-3}$.

For starburst galaxies, it is expected to have a continuous star formation over at least a galactic dynamical timescale, then, the assumption of a continuous rather than an instantaneous burst of star formation would be more accurate. We generated the PEGASE grid of Kew01 for a continuous starburst model and found, as expected, that this corresponds to SF and composite objects (see Fig. 4b), since this is the limit used by Kew01 to parametrize an extreme starburst line in the BPT diagrams.

The photoionization grids generated with the STARBURST99 code were not hard enough  to produce the needed [{S\,\textsc{ii}}] flux to enclose all galaxies in the S2N2 diagram. The BPT diagrams are most sensitive to the spectral index of the ionizing radiation field in the 1-4 ryd interval, and the PEGASE ionizing stellar continuum is harder in this range than that of STARBURST99, being PEGASE the only models that encompass nearly all of the observed starburst on all three of the BPT diagrams. We also tried the grids of Levesque et al. (2010), which use the STARBURST99 code, but these grids comprise an insufficiently hard ionizing radiation field, leading to deficiencies in the [{S\,\textsc{ii}}] fluxes produced by the models.

In galaxy surveys there are at least three methods commonly used to deal with the presence of AGNs. The first one consists on removing galaxies hosting AGNs by cross-correlating the sample with published AGN catalogues (e.g. Condon et al. 2002, Serjeant et al. 2002). The second one deals with the identification of the galaxies through the so called BPT diagram. Nevertheless, at z $\gtrsim$ 0.5, the {H$\alpha$} line is redshifted out of the optical range. The third method consists in subtracting AGNs in a statistical manner, used when no other methods are applicable (e.g. Tresse $\&$ Maddox 1998).

In addition to those methods, we propose the use of the S2N2 diagram, which has demonstrated the ability of accurately segregate SF from composite and AGN galaxies. The S2N2 diagram has the following advantages: it is not necessary an extinction correction, since all the emission lines are close in wavelength; requires only a small spectral range, making it suitable for surveys of limited spectral coverage; the SF and composite divisions of the S2N2 diagram offer less contamination in all cases, with respectively higher number of galaxies for SF and composites than using only the N2 ratio; additionally, the user can choose any of the SF galaxies divisions provided for the S2N2 diagram, either if the smallest contamination from composite galaxies (Eq. 1), or selecting the most SF galaxies (Eq. 2) is required. Finally, the use of the [{S\,\textsc{ii}}] lines do not reduce the number of galaxies, since those lines for AGN galaxies are, in mean $\sim$1.3 times stronger than the [{O\,\textsc{iii}}] line used in the BPT diagrams. Then, comparing the number of galaxies of all the types, the S2N2 diagram have 1$\%$ more galaxies than the log([{N\,\textsc{ii}}] /{H$\alpha$}) vs log([{O\,\textsc{iii}}] /{H$\beta$}) diagram. Although this diagram has been used in the past for galaxies, it is the first time that it is presented as a diagnostic diagram for classifying galaxies. Given its advantages, we propose the use of this diagram as an alternative to the BPT diagrams and N2 ratio to classify galaxies.

\section{Evolutionary effects on the BPT diagrams}

As explained in Section 3, the BPT and other optical emission lines diagnostic diagrams have become important in the classification of galaxies. In this section, our aim is to investigate the effects of the evolution of galaxies from the three BPT diagrams. For this purpose, and with the objective of increase our number of galaxies, we did not take any restriction in magnitude, as detailed in the sample selection.

\begin{figure*}[ht!]
\centering
\includegraphics[scale=0.7]{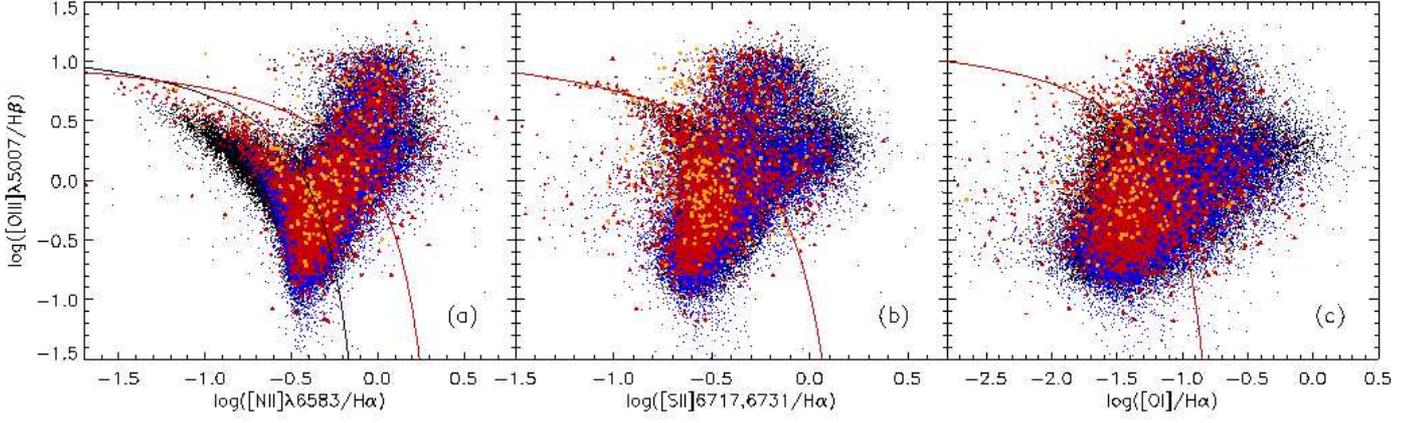}
\caption{Evolution of line ratios from the BPT diagrams. Black dots show galaxies in the redshift range $z_0$, blue dots galaxies in the $z_1$ range, red big triangles represent galaxies in the completeness of the $z_2$ range, while small red triangles galaxies out of the completeness,  yellow big circles represents galaxies in the completeness of the $z_3$ range, and small circles galaxies out of the completeness. The black solid line shows the Kauf03 limit for SF galaxies, and the red solid line shows the Kew01 limit for starburst galaxies in the three BPT diagrams.}
\end{figure*}

\begin{figure*}[ht!]
\centering
\includegraphics[scale=0.7]{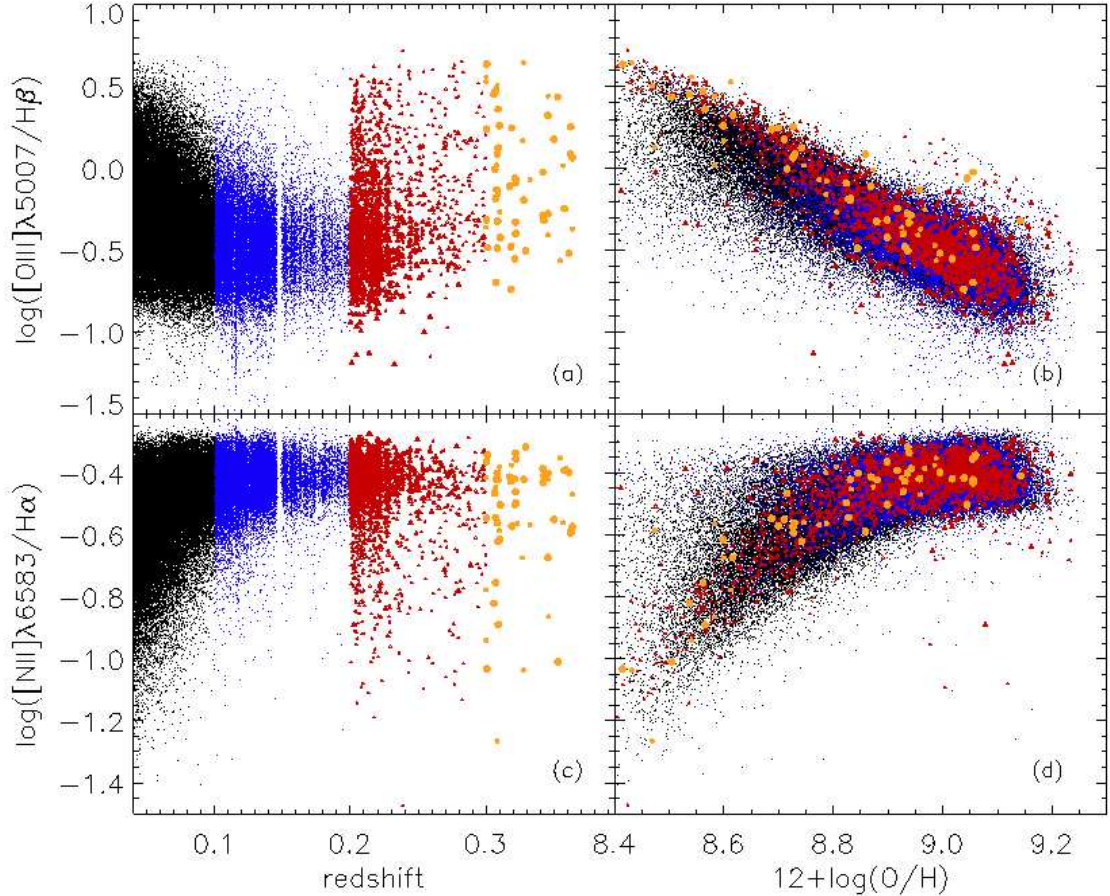}
\caption{Evolution with redshift of the log([{O\,\textsc{iii}}] $\lambda$5007/{H$\beta$}) ratio for star--forming galaxies (left), and 12+log(O/H) vs log([{O\,\textsc{iii}}] $\lambda$5007/{H$\beta$}) for the same galaxies (right). Symbols are the same as in Fig. 5. The gap observed at $z \sim 0.145$ is due to the 5577 {\AA} sky line falling in the {H$\beta$} line, missing galaxies around this redshift.}
\end{figure*}

In Fig. 5 we show the three BPT diagrams for the four redshift samples. As redshift increases, we observe that [{O\,\textsc{iii}}] $\lambda$5007/{H$\beta$} goes toward higher values. In order to explain this shift, in Fig. 6 we plotted the ratio [{O\,\textsc{iii}}] $\lambda$5007/{H$\beta$} versus redshift and metallicity only for SF galaxies selected with the Kauf03 criterion. The gap observed around $z\sim$ 0.145 (see Fig. 6a, c) is due to the {H$\beta$} line falling nearby the 5577 {\AA} sky line, because the residuals are significant and, as a consequence, measurements of {H$\beta$} around this redshift were lost. As shown in Fig. 6b, there is a clear tendency of the [{O\,\textsc{iii}}] $\lambda$5007/{H$\beta$} ratio towards higher values with redshift, which is explained by examining the same ratio against 12+log(O/H). The ratio [{O\,\textsc{iii}}] $\lambda$5007/{H$\beta$} has demonstrated to correlate linearly with metallicity (see, for example, Liang et al. 2006). Then, a decrement in 12+log(O/H) will result in higher values of [{O\,\textsc{iii}}] $\lambda$5007/{H$\beta$} (see Fig. 6b). We observe a decrement of $\sim$0.2 dex in  [{O\,\textsc{iii}}] $\lambda$5007/{H$\beta$}, and a decrement of $\sim$0.1 dex in 12+log(O/H) for the $z_3$ redshift range with respect to the $z_0$ range.

In previous papers (Lara-L\'opez et al. 2009a, b) we reported a decrement in 12+log(O/H) of $\sim$0.1 dex for the redshift range $0.3 < z < 0.4$  comparing galaxies in the same range of luminosity at different redshift intervals. Since the possible bias, such as luminosity, mass and aperture effects of those samples were carefully studied, we demonstrated there that this decrement in metallicity is due to an intrinsic evolution of the galaxies.

Although our $z_3$ sample corresponds to luminous galaxies, if we compare galaxies with the same luminosity, taking as a reference our previous papers, the metallicity decrement will be again of $\sim$0.1 dex, and as consequence, the effects on the BPT diagrams will be the same. Therefore, the evolution observed in the  [{O\,\textsc{iii}}] $\lambda$5007/{H$\beta$} lines ratio toward higher values in the three BPT diagrams, could be attributed to a metallicity evolution.

On the other hand, we analyze the  [{N\,\textsc{ii}}]/{H$\alpha$} ratio against redshift and metallicity (see Fig.6c, d). The [{N\,\textsc{ii}}]/{H$\alpha$} ratio is also a metallicity index, commonly known as N2, and it has been widely studied since it is not severely affected by dust extinction (see Pettini $\&$ Pagel 2004). Among the calibrations of the N2 index, we have for example those of Raimann et al. (2000), Denicol\'o et al. (2002), and Liang et al. (2006). In Fig. 6d we observe a clear increasing trend of metallicity following the increase of the N2 index up to 12+log(O/H) $\sim$9.0. The galaxies with 12+log(O/H) $>$9  show a flattening and a slightly decrease of the N2 index with metallicity (see Fig. 6d). This trend was explained by Kewley et al. (2002) using photoionization models as follows: when the secondary production of nitrogen dominates, at somewhat higher metallicity, the [{N\,\textsc{ii}}]/{H$\alpha$} line ratio continues to increase, despite the decreasing electron temperature. Eventually, at still higher metallicities, nitrogen becomes the dominant coolant in the nebula, and the electron temperature falls sufficiently to ensure that the nitrogen line weakens with increasing metallicity. Liang et al. (2006), using SDSS galaxies with redshifts $0.04 < z < 0.25$, observed a small decrement of the N2 index against metallicity; this turnover of the N2 index is more evident for the higher redshifts $z_2$ and $z_3$ in our sample (see Fig. 6d). The turnover of the N2 ratio will produce in the [{O\,\textsc{iii}}] $\lambda$5007/{H$\beta$} vs.  [{N\,\textsc{ii}}]/{H$\alpha$} BPT diagram, the turnover zone around N2 $\sim-0.4$, which is more evident in a density plot (see Fig. 2).

Regarding the two left BPT diagrams of Fig. 5b and c, since they share the ratio [{O\,\textsc{iii}}] $\lambda$5007/{H$\beta$}, the evolutionary effects due to a decrement in metallicity will be the same as discussed above. The ratio [{S\,\textsc{ii}}]/{H$\alpha$} has never been used before as a metallicity indicator because it is far more sensitive to ionization than to metallicity (Liang et al. 2006). Moreover, it is double--valued with metallicity (see Fig. 4), whereas the ratio [OI]/{H$\alpha$} is not a metallicity indicator.

Therefore, after analyzing all the ratios involved in the three BPT diagrams, we concluded that the evolution of galaxies in the three BPT diagrams is shown through the [{O\,\textsc{iii}}] $\lambda$5007/{H$\beta$} ratio. Since this ratio is a metallicity indicator, any decrement in metallicity will result in higher values of the [{O\,\textsc{iii}}] $\lambda$5007/{H$\beta$} ratio.

%Hereafter, when comparing all the redshift intervals will be important to take into account that the magnitude completeness of the redshift interval $z_3$ corresponds to higher magnitudes than the lower redshift intervals (see Fig. 1). Nevertheless, the metallicity decrement found in this section is the same that the one found in our previous papers, in which we compare galaxies at the same luminosities

\section{Evolution of the mass--metallicity and luminosity--metallicity relations}

It has been demonstrated that the metallicity and mass of SF galaxies are strongly correlated, with massive and luminous galaxies showing higher metallicities than less massive galaxies (see Sect. 1).

The masses of our galaxies were estimated using the STARLIGHT code, which fit an observed spectrum with a combination of 150 SSPs from the evolutionary synthesis models of Bruzual $\&$ Charlot (2003), computed using a Chabrier (2003)  initial mass function between 0.1 and 100 M$_{\odot}$, and $``$Padova 1994" evolutionary tracks. The 150 base elements span 25 ages between 1 Myr and 18 Gyr, and six metallicities from $Z=0.005$ to 2.5 $Z_{\odot}$. As argued by Mateus et al. (2006), the inclusion of very low $Z$ SSPs in the base inevitably leads to larger stellar masses. A comparison with the Kauffmann et al. (2003b) mass estimates, which are based on a library of model galaxies constructed with $Z >$ 0.25 $Z_{\odot}$, results in systematic discrepancies of about 0.1 dex (for details see Mateus et al. 2006). The masses of our galaxies were corrected for aperture effects based on the differences between the total galaxy magnitude in the $r$ band, and the magnitude inside the fiber, assuming that the mass--to--light ratio does not depend on the radius (see Mateus et al. 2006 for details).

A histogram of our mass estimates is shown in Fig. 7, where a larger fraction of massive galaxies are observed at highest redshifts. In Fig. 8 we derived the $M-Z$ and $L-Z$ relations for the galaxies of our sample. As explained  in previous sections, galaxies of the $z_0$ and $z_1$ samples are complete in luminosity, while for the $z_2$ and $z_3$ samples, the completeness criterium is not taken into account.

The $M-Z$ relation of T04, which is valid over the range 8.5 $<$ log(M$_{star}$/M$_{\odot}$) $<$ 11.5, shows a steep $M-Z$ relation for masses from $10^{8.5}$ to $10^{10.5}$ M$_{\odot}$ that flattens at higher masses. In such study, T04 analyzed galaxies with redshift ranges $0.005 < z < 0.3$. It is important to notice that in our $M-Z$ relation, a flatness is not observed for masses $\gtrsim$ $10^{10.5}$ for the redshift range $z_0$, (see Fig. 8a), but this flatness is observed for the higher redshift samples. Then, the flatness observed by T04 depends on the redshift range observed. In order to establish the bias-free $M-Z$ relation for local galaxies, Kewley $\&$ Ellison et al. (2008) recalibrated the $M-Z$ relation of T04 with galaxies at 0.04 $<$ z $<$ 0.1, since 0.04 is the minimum redshift to avoid fibers effects (Kewley et al. 2005).

In Fig. 8 ($M-Z$ and $L-Z$ relations) the metallicity decrement for the $z_3$ redshift sample discussed in our previous articles (Lara-L\'opez et al. 2009a, b) is also evident. In order to fit our local $M-Z$ relation, we estimated the mode of the metallicity of the galaxies in mass bins of 0.1 dex and fit them with a second order polynomial (y=$a_0+a_1x+a_2x^2$), with $a_0=-0.467, a_1=1.611 , a_2=-0.067$. We also fit a second order polynomial to our $M-Z$ relation for galaxies at $z_3$, with $a_0=-0.632, a_1=1.557, a_2=-0.063$. All the fits are shown in Fig. 9. According to them, our $M-Z$ relation for the galaxies at $z_3$ is $\sim$ 0.2 dex lower compared to our local galaxy sample. Additionally, in Fig. 9 we compare the $M-Z$ fits from literature at different redshifts with our results. At $z \sim 0.15$ we represent the calibration of T04, at $z \sim 0.07$ the T04 recalibration of  Kewley $\&$ Ellison et al. (2008), as well as our fit for the local ($z_0$) $M-Z$ relation. At higher redshift, we represent our fit to the $M-Z$ relation for galaxies at $z_3$, as well as the fit of Erb et al. (2006a)  at z $\sim$2.2  scaled to the T04 metallicity calibration. Due to their high redshift, Erb et al. (2006a) used the N2 method and the calibration of Pettini $\&$ Pagel (2004) to estimate their metallicities. We converted their N2 metallicities to the R$_{23}$ calibration of T04 with the metallicity conversions given in Kewley $\&$ Ellison et al. (2008). Even with the dispersion of our local sample, our  $M-Z$ fit is a little lower, but in a good agreement with those of T04 and Kewley $\&$ Ellison et al. (2008). Since we are using the T04 calibration of the R$_{23}$ method to estimate metallicities, the main differences with the fit of T04 are the redshift ranges, as discussed above, and the mass estimates, since T04 and Kewley $\&$ Ellison et al. (2008) adopted a Kroupa et al. (2001) IMF, while we are using a Chabrier (2003) IMF.

Although redshift ranges are different, the comparison of our fit for the $z_3$ sample with the Erb et al. (2006a) data at z$\sim$2.2, which also use a Chabrier IMF, are similar in 12+log(O/H) (see Fig. 9). As will be explained in the next sections, our $z_3$ sample is conformed mainly by spiral galaxies, while the Erb et al. (2006a) sample corresponds to a mix of morphological types. A possible explanation for the high metallicities of Erb et al. (2006a), or to the lower metallicities of our sample, is given by Calura et al. (2009) who, using models that distinguish among different morphological types through the use of different infall, outflow and star formation, reproduce the mass-metallicity relation in galaxies of all morphological types, taking as a reference the observational $M-Z$ relations of Kewley $\&$ Ellison (2008), Savaglio et al. (2005), Erb et al. (2006a), and Maiolino et al. (2008). In his work Calura et al. (2009) predicts that at any redshift, elliptical galaxies will present the highest stellar masses and the highest metallicities, whereas the irregulars are the least massive and metallic galaxies, being spiral galaxies at an intermediate stage.

This means that, being our $z_3$ sample composed only by spiral galaxies, our metallicities will be lower than if our sample were composed by a mix of morphological types. The observed metallicities and SFRs for the Erb et al. (2006a) sample at z$\sim$2.2, according to the study of Calura et al. (2009), indicate that their galaxies are likely to represent a morphological mix, partly composed of spirals (or proto-spirals) and partly of ellipticals (or proto-ellipticals). Calura et al. (2009) predicts for the Erb et al. (2006a) galaxies at z$\sim$2.2  lower metallicities ($\sim$0.3 dex) if the sample were composed by spiral galaxies. Then, the similarities in 12+log(O/H) of our $M-Z$ relation at $z_3$ with the one of Erb et al. (2006a) at z$\sim$2.2, could be explained by the morphological selection in each case: lower metallicities in our $M-Z$ relation can be addressed to the prominence of spiral galaxies, whereas larger metallicities in the $M-Z$ relation of the Erb et al. (2006a) is a consequence of a sample formed by a mix of morphological types. In other words, the morphology of the galaxies is crucial in deriving and comparing the metallicity and the $M-Z$ relation.

An additional point to take into account, is that our samples plotted in Fig 9 are selected with differente magnitude completeness, then, our $z_3$ redshift sample is high luminous and massive than our local one. This must be taken into account when comparing both $M-Z$ relations. Unfortunately, our $z_0$ sample does not have a significant number of galaxies in the same absolute magnitude range of the $z_3$ sample, making imposible to generate a local $M-Z$ relation comparable in luminosity to the $z_3$ one.

Comparing our $M-Z$ relation with that of high-$z$ samples in the literature, we have the $M-Z$ relation of Savaglio et al. (2005) at $z \sim 0.7$. In their study, they found only a slightly decrement in metallicity for galaxies at $z \sim 0.7$ compared with the local one of T04, which is inconsistent with the $\sim 0.2$ dex decrement found in our $z_3$ $M-Z$ relation. Nevetheless, as pointed out by Rodrigues et al. (2008), the sample of Savaglio et al. (2005) have spectra with low $S/N$ and spectral resolution, as well as extinction problems. With a more consistent result, Rodrigues et al. (2008) generated the $M-Z$ relation for galaxies at $z \sim 0.7$, finding a decrement in metallicity of $\sim$0.3 dex compared with the local one of T04. Unfortunately, given the small range in their stellar masses, it was not possible to constrain the evolution of the shape of their $M-Z$ relation.

\begin{figure}[ht!]
\centering
\includegraphics[scale=0.80]{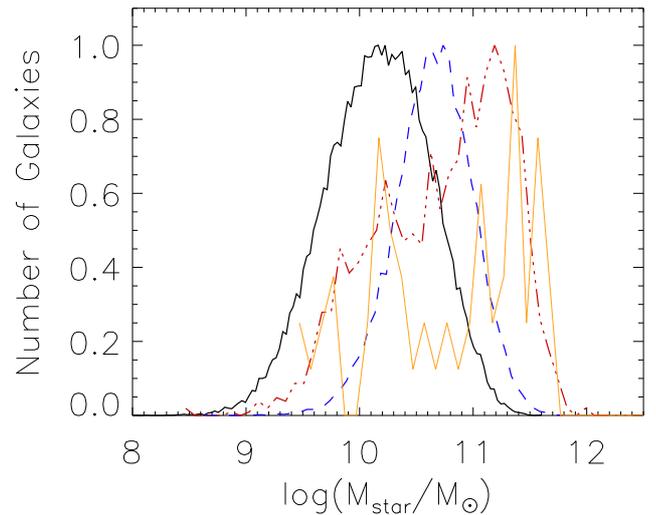}
\caption{Normalized mass histogram for all our samples. Dark solid line represent galaxies at $z_0$, dashed line galaxies at $z_1$, point dashed line galaxies at $z_3$, and clear solid line galaxies at $z_3$. The histograms were constructed do not taking into account the completeness for galaxies at $z_2$ and $z_3$.}
\end{figure}

There are two main ways to explain the origin of the $M-Z$ relation. The first one is related to the well-known effect of downsizing (e.g. Cowie et al. 1996, Gavazzi $\&$ Scodeggio 1996), in which lower mass galaxies form their stars later and on longer time-scales than more massive systems, implying low star formation efficiencies in low--mass galaxies (Efstathiou 2000; Brooks et al. 2007; Mouhcine et al. 2008; Tassis et al. 2008; Scannapieco et al. 2008, Ellison et al. 2008). Therefore, low--mass galaxies are expected to show lower metallicities. Supporting this scenario, Calura et al. (2009) reproduced the $M-Z$ relation with chemical evolution models for ellipticals, spirals and irregular galaxies, by means of an increasing efficiency of star formation with mass in galaxies of all morphological types, without the need for outflows favoring the loss of metals in the less massive galaxies. In a recent study that supports this result for massive galaxies, Vale Asari et al. (2009), model the time evolution of stellar metallicity using a closed-box chemical evolution picture. They suggest that the $M-Z$ relation for galaxies in the mass range from $10^{9.8}$ to $10^{11.65}$ M$_{\odot}$ is mainly driven by the star formation history and not by inflows or outflows.

\begin{figure*}[ht!]
\centering
\includegraphics[scale=0.90]{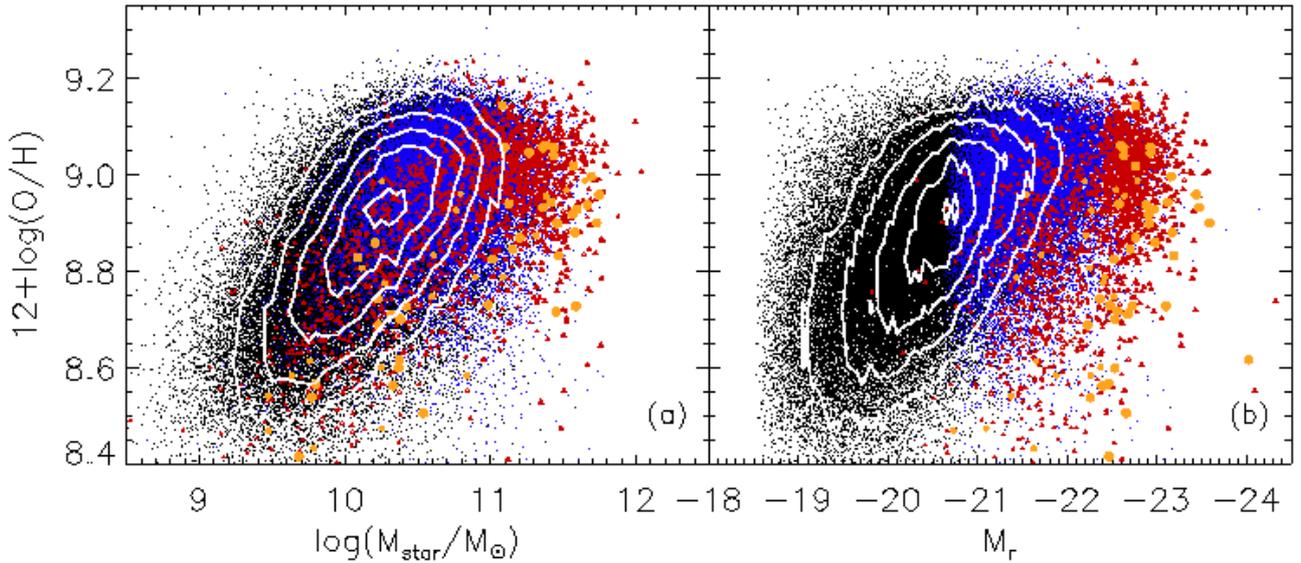}% SFRParaGalaxiasHII-BPT-TodosZs.pro (ese programa la  hace)
\caption{ (a) Relation between the stellar mass and 12+log(O/H) ($M-Z$ relation). (b) Relation between the absolute Petrosian $r$ magnitude and 12+log(O/H) ($L-Z$ relation) for our sample of galaxies. The cut observed in Fig. (b) for the $z_1$ sample, is due to the 5577 {\AA} sky line (see the text). In both relations, white contours represent, from outside to inside, 15, 30, 50, 70, and 90$\%$ of the maximun density value of the $z_0$ redshift sample (black dots), and are plotted only as a visual aid. Colors and symbols follow the same code used in Fig. 5.}
\end{figure*}

A second scenario to explain the $M-Z$ relation is attributed to metal and baryon loss due to gas outflow, where low--mass galaxies eject large amounts of metal--enriched gas by supernovae winds before high metallicities are reached, while massive galaxies have deeper gravitational potentials which helps to retain their gas, thus reaching higher metallicities (Larson 1974; Dekel $\&$ Silk 1986; MacLow $\&$ Ferrara 1999; Maier et. al. 2004; T04; De Lucia et al. 2004; Kobayashi et al. 2007; Finlantor $\&$ Dave 2008). As pointed out in the high--resolution simulations of Brooks et al. (2007), supernovae feedback plays a crucial role in lowering the star formation efficiency in low--mass galaxies. Without energy injection from supernovae to regulate the star formation, gas that remains in galaxies rapidly cools, forms stars, and increases its metallicity too early, producing a $M-Z$ relation too flat compared to observations.

\begin{figure}[ht!]
\centering
\includegraphics[scale=0.80]{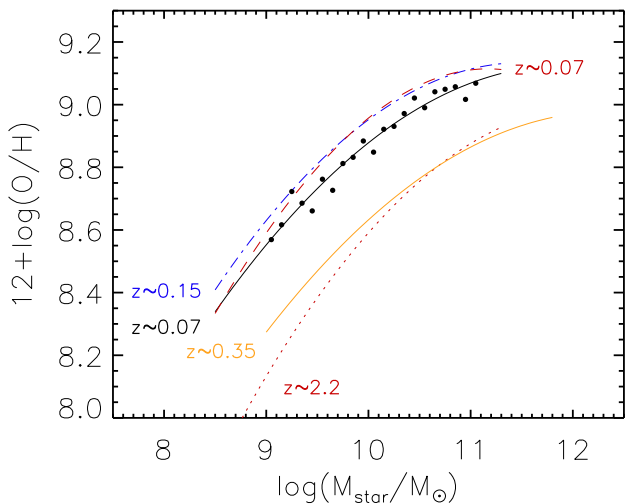}
\caption{Evolution of the mass-metallicity relation observed at different redshifts. Point-dashed line represents the curve of Tremonti et al. (2004) at z$\sim$0.15, dashed line represents the T04 recalibration of Kewley $\&$ Ellison (2008) at z$\sim$0.07, circles are the mode metallicity in log(M$_{star}$/M$_{\odot}$) bins of 0.1 for our $z_0$ sample, solid dark and clear curve represents our fit for $z_0$ (mode bins) and $z_3$, respectively. Dotted line represents the fit of Erb et al. (2006a) at z$\sim$2.2.}
\end{figure}

An additional  interpretation of the $M-Z$ relation is linked to some properties of star formation, as the IMF. K\"oppen et al. (2007) suggested that the $M-Z$ relation can be explained by a higher upper-mass cutoff in the IMF in more massive galaxies.

Finally, we also generated the $L-Z$ relation for our redshift samples (see Fig. 8b). Nevertheless, our $z_3$ sample is restricted to a small range in luminosity, making impossible to fit a curve. The local $L-Z$ relation is well established by e.g. T04, then, due to our small luminosity range  at $z_3$ we can not conclude anything about the $L-Z$ relation. Nevertheless, the $M-Z$ relation has demonstrated to be more stronger and tighter than the $L-Z$ relation, confirming that stellar mass is a more meaningful physical parameter than luminosity when both are compared with gas metallicity (Savaglio et al. 2005).

\begin{figure}[ht!]
\centering
\includegraphics[scale=0.80]{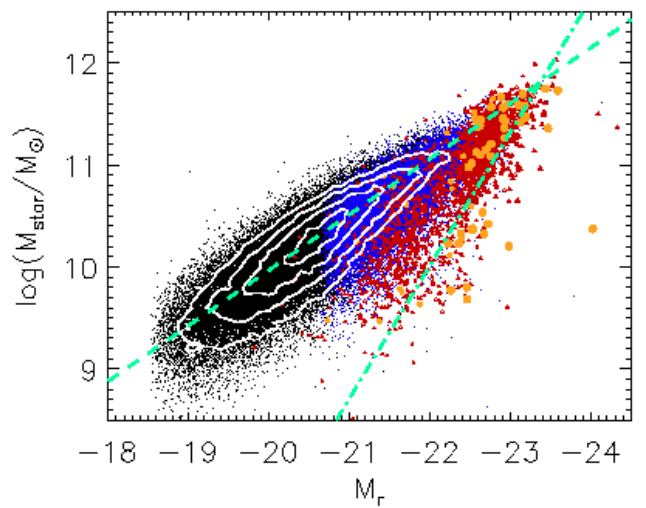}
\caption{Mass versus Petrosian absolute k-corrected magnitude for all our redshift samples. White contours represent, from outside to inside, 10, 25, 50, and 80$\%$ of the maximun density value of the $z_0$ sample. Dashed line represents the fit to the $z_0$ sample, while point-dashed line the fit to the $z_3$ sample. Colors and symbols follow the same code used in Fig. 5.}
\end{figure}

\subsection{Evolution of the mass-to-light ratio}

As explained by Erb et al. (2006a), at higher redshifts, the $M-Z$ relation is clearly more physically meaningful than the $L-Z$ relation. A corollary is that the local $L-Z$ relation is simply a result of the strong  correlation between mass and luminosity at low redshift.

We also analyzed the evolution of the mass-to-light ($M/L$) ratio (see Fig. 10). For a given mass, we observed higher luminosities for the $z_3$ sample compared with the local one, which means lower $M/L$ ratios as redshift increase. In order to observe this evolution, we fit a line ($y=a_0+a_1x$) to the $z_0$ and $z_3$ redshift samples (see Fig. 10), obtaining $a_0=-0.924$ and  $a_1=-0.544$ for $z_0$, and $a_0=-18.686$ and  $a_1=-1.304$ for $z_3$. The variation in $M/L$ at a given rest frame optical luminosity can be as much as a factor of $\sim$70 (Shapley et al. 2005), which means that for any range in luminosity there exist an extended range of stellar masses. This large variation in $M/L$ explains the lack of correlation in the $L-Z$ relation for the $z_3$ sample compared to the local relation. For a small range of absolute magnitudes in the $z_3$ sample, we have a widely range of masses, making possible to generate a $M-Z$ relation. At higher redshifts the effect is the same, as pointed out by Erb et al. (2006a), finding for star-forming galaxies at z$\sim$2.2 in a small range of luminosity, a wide range of stellar mass.

\section{Morphology indicators and SFR}

The variation of SFR activity and young stellar content along the Hubble sequence is one of the most recognizable features of galaxies. In fact, this variation in stellar content is part of the basis of the Hubble classification itself (Hubble 1926), and understanding its physical nature and origin is fundamental to comprehend the galaxy evolution (for a review see Kennicutt 1998 and Kennicutt et al. 1994). The general picture, presented by Roberts (1963), Searle et al. (1973), Larson $\&$ Tinsley (1978), and Kennicutt et al. (1994), point that early--type galaxies (types S0-Sb) represent systems which formed most of their gas into stars on timescales much less than the Hubble time, while the disks of late--type systems (Sc--Im) have formed stars at roughly a constant rate since they formed.

For most of the 20th century, catalogues of morphologically classified galaxies were compiled by individuals or small teams of astronomers (e.g. Sandage 1961, de Vaucouleurs 1991). Nowadays selection criteria is based on galaxy properties such as color, concentration index, spectral features, surface brightness profile, other structural parameters, or some combination of these (e.g. Strateva et al. 2001, Abraham et al. 2003, Kauffmann et al. 2004, Conselice 2006, Scarlata et al. 2007). With the advent of modern surveys, such as the SDSS and with the participation of thousands of volunteers, it was possible the development of the Galaxy Zoo project (e.g. Lintott et al. 2008), providing visual morphological classification for more than $10^7$ galaxies.

In this section, we will focus on the relations at different redshifts between the SFR, metallicity, mass, and morphology of SF  galaxies selected with the Kauffmann et al. (2001) criteria, as explained in the Section 2.

\subsection{Evolution of the SFR}

We estimate the  SFR with the {H$\alpha$} emission line flux following the Kennicutt (1998) expression:

\begin{equation}
{\rm SFR}[{{\rm M}_{\odot}}\;{\rm yr}^{-1}]=7.9 \;{\rm x}\; 10^{-42}{\rm L ({\rm H\alpha}})\; [{\rm ergs}\;{\rm s}^{-1}],
\end{equation} where ${\rm L ({\rm H\alpha}})$ denotes the intrinsic {H$\alpha$}  luminosity, and {H$\alpha$} is corrected by dust extinction and underlying stellar absorption as explained in Sec. 2. This calibration is derived from evolutionary synthesis models that assume solar metallicity and no dust, and  is valid for a T$_e$=10$^{4}$ K and case B recombination

Recent studies have explored the relationship between the stellar mass and the SFR in galaxies at different redshifts. It has been shown that SFR critically depends on the galaxy mass both at low and high redshifts (e.g. Gavazzi et al. 2002, Brinchmann et al. 2004, Dickinson et al. 2004, Feulner et al. 2005, Papovich et al. 2006).

In our sample, galaxies with high SFRs are more abundant at higher redshifts (see Fig. 11), a fact already observed in non biased samples (e. g. Noeske et al. 2007a). In Fig. 11a, we show 12+log(O/H) against log(SFR). Note that, although our $z_3$ sample of galaxies is biased to the most luminous and massive galaxies, the observed decrement of $\sim$ 0.1 dex in 12+log(O/H) found in Lara-L\'opez et al. (2009a, b) is also present. Regarding the $z_0$ sample of galaxies, there is a clear sequence with galaxies going toward higher values of SFR as metallicity increases. This tendency can be explained from the $z_0$ sample in the $M-Z$ relation of Fig. 8, where massive galaxies corresponds to the highest metallicity galaxies, and for more massive galaxies, we expect higher SFRs (see Fig. 11). Also, we can slightly appreciate a population of galaxies with higher SFR (see Fig. 11a). This population will form a $tail$ when the mass is taken into account, as will be shown in Fig. 11b. As redshift increases, we appreciate in Fig. 11a for galaxies at $z_2$ and $z_3$, a flattening of the SFR vs. 12+log(O/H) relation, with most of the galaxies showing log(SFR) between 1 and 2.

In Fig. 11b, we show the log(M$_{star}$/M$_{\odot}$) versus log(SFR) plot. Galaxies at $z_0$ show a main sequence, where massive galaxies have higher SFRs. This main sequence was identified by Noeske et al. (2007a), studying galaxies with redshifts from 0.2 to 1.1, finding that this main sequence moves as a whole to higher SFR as redshift increases. The SSFR, defined as the total SFR divided by the stellar mass, reflects the strength of the current burst of star formation relative to the underlying galaxy mass. Deep galaxy surveys have consistently found that the SSFR depends strongly on both ${\rm M}_{\odot}$ and redshift, with the bulk of star formation occurring earlier in massive galaxies than in less massive systems (Guzm\'an et al. 1997, Brinchmann $\&$ Ellis 2000, Juneau et al. 2005, Bauer et al. 2005, Bell et al. 2005, P\'erez-Gonz\'alez et al. 2005, Feulner et al. 2005, Papovich et al. 2006, Caputi et al. 2006, Reddy et al. 2006).

We also analyzed the evolution of the SSFR as a function of the metallicity and the stellar mass. Interestingly, in the 12+log(O/H) vs. SSFR diagram (Fig. 12a), the observed population of Fig. 11a at $z_0$ is more evident, showing a higher SSFR (SSFR $>-10$) than the other galaxies at the same redshift. We are going to investigate this $tail$ in more detail in the next subsection.

In Fig. 12b, we show the log(M$_{star}$/M$_{\odot}$) versus log(SSFR) plot. The SSFR increase with redshift, showing  for more massive galaxies a tendency toward lower SSFR values, which is in agreement with the results of Noeske et al. (2007b), for galaxies with z $>$ 0.2. Massive galaxies shows lower SSFR because they probably have low gas fractions and have thus nearly finished assembling their stellar mass (Erb et al. 2006b, Reddy et al. 2006). On the other hand, the presence of dust has demonstrated to play an important role deriving the SSFR, as shows by Pannella et al. (2009), getting for dust free galaxies a flat SSFR for redshift bins centered at $z \sim 1.6$ and 2.1, instead of a drop with increasing mass. In our samples, the SFR and SSFR were derived through the dust corrected  {H$\alpha$} flux, showing that for redshifts less than 0.4, this flatness is absent.

We found also evidences of two populations in Fig. 12b, one at redshift $z_0$ which concentrate in a square delimited by the contour plots, and another one showing SSFR $>-10$, which is more evident for the higher redshifts samples, but it is also evident in the $z_0$ sample. This population of galaxies will be explained in the next subsection as a consequence of the morphology of the galaxies.

\begin{figure*}[ht!]
\centering
\includegraphics[scale=0.90]{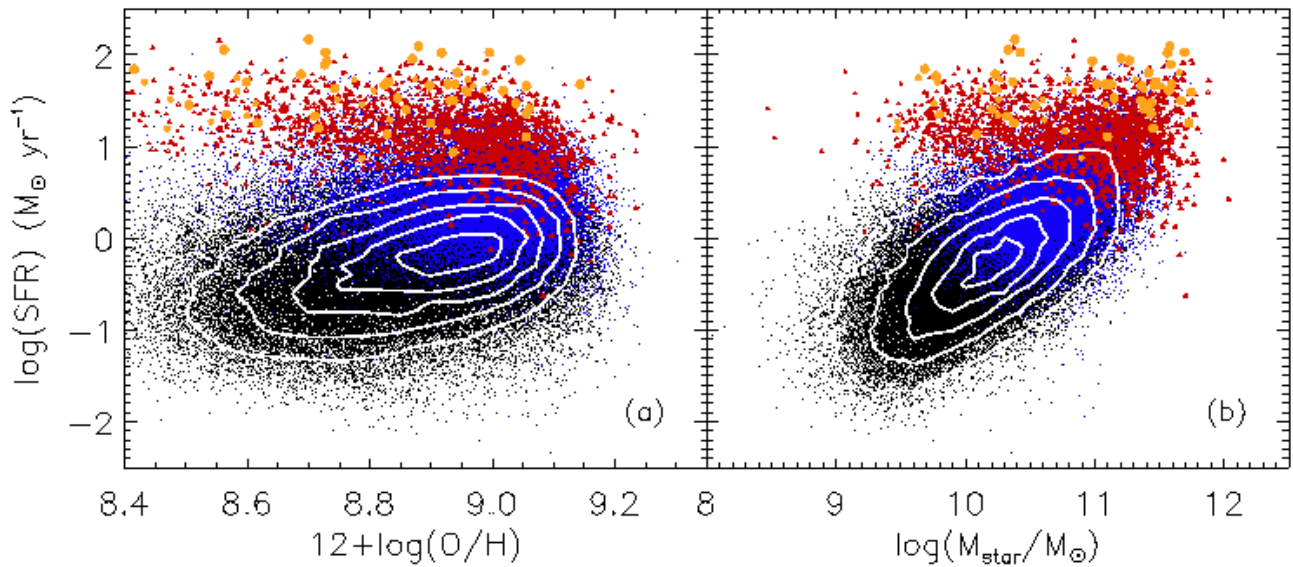}
\caption{Metallicity and mass versus log(SFR), contours correspond to the $z_0$ sample in both plots. White contours represent, from outside to inside in panel $a$: 15, 30, 50, 70, and 90$\%$, and in panel $b$: 5, 15, 35, 65, and 85$\%$ of the maximun density value of the $z_0$ sample. Colors and symbols follow the same code used in Fig. 5.}
\end{figure*}

\begin{figure*}[ht!]
\centering
\includegraphics[scale=0.90]{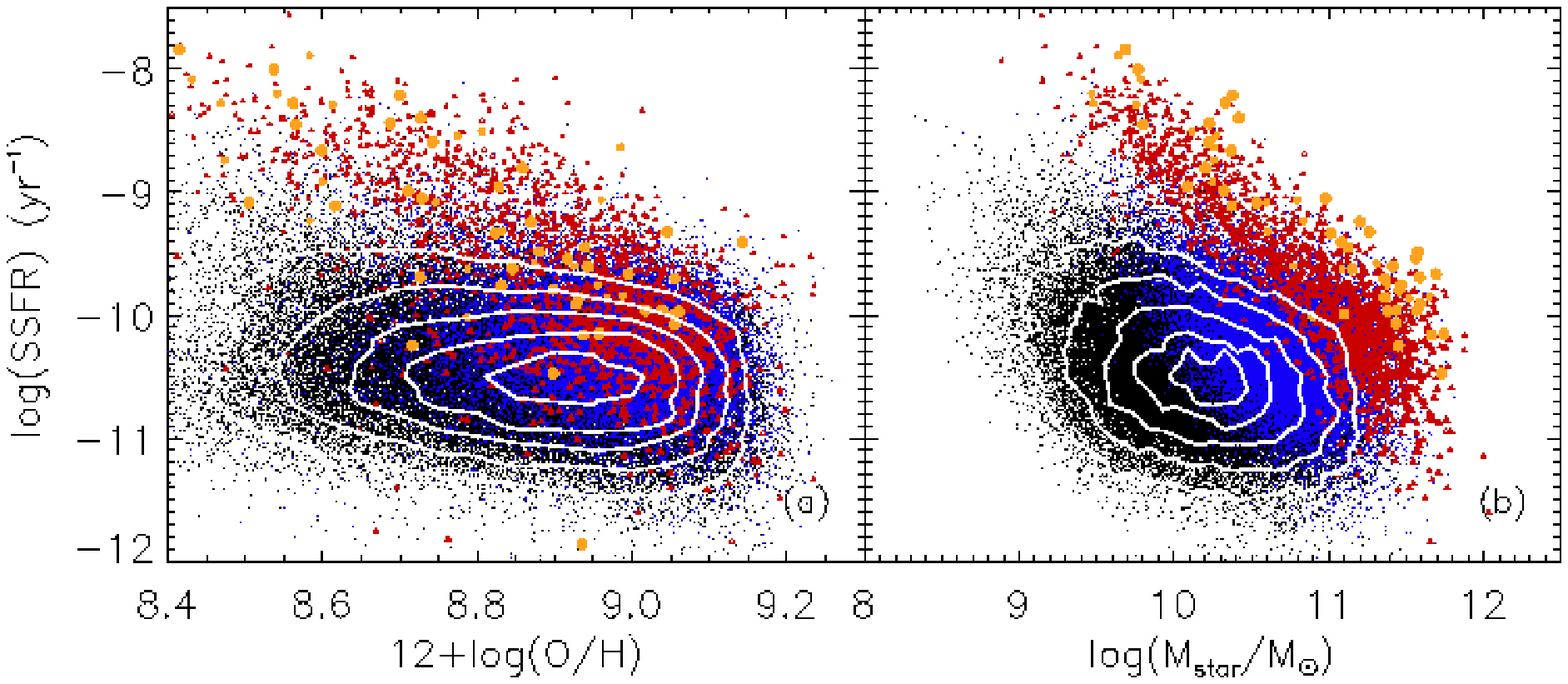}
\caption{Metallicity and mass versus log(SSFR), contours correspond to the $z_0$ sample in both plots. White contours represent, from outside to inside in panel $a$: 10, 20, 40, 60, and 90$\%$, and in panel $b$: 5, 15, 35, 60, and 80$\%$ of the maximun density value of the $z_0$ sample. Colors and symbols follow the same code used in Fig. 5.}
\end{figure*}

\subsection{The SSFR as a morphology indicator}

\begin{figure*}[ht!]
\centering
\includegraphics[scale=0.7]{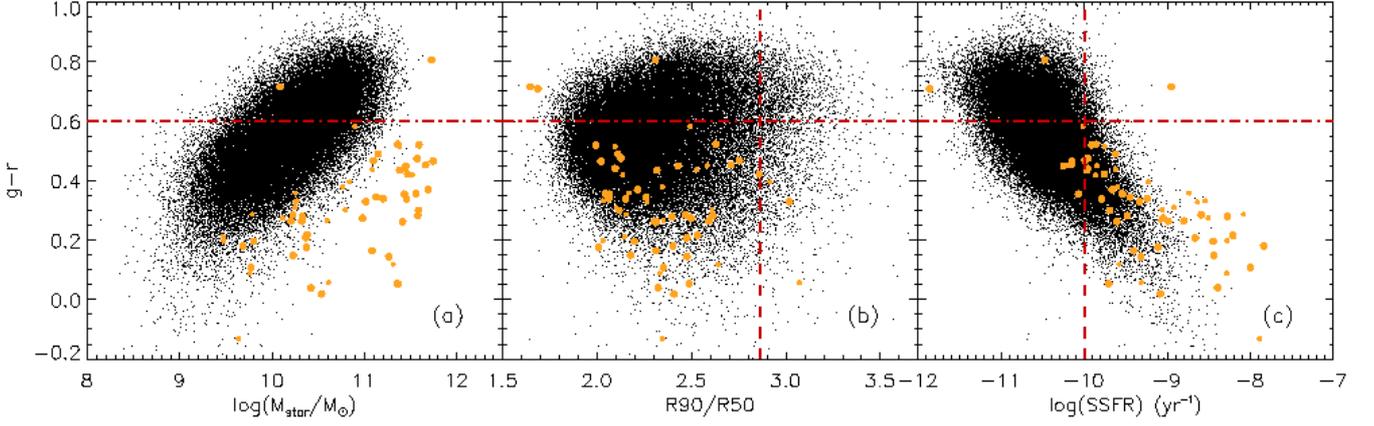}   %esta grafica la hace ColoresHIITodasMagnitudesConMetalicidades.pro
\caption{Principal morphological classificators. In panel a, b, and c, we show the log(M$_{star}$/M$_{\odot}$), concentration index  c=R$_{90}$/R$_{50}$, and log(SSFR) versus $g-r$ color, respectively. Black dots and yellow circles represent galaxies at $z_0$ and $z_3$, respectively. In each panel the dashed line shows the standard limit to segregate early from late--type galaxies, with the $z_3$ galaxies concentrated in the late--type region in each panel.}
\end{figure*}

\begin{figure*}[ht!]
\centering
\includegraphics[scale=0.7]{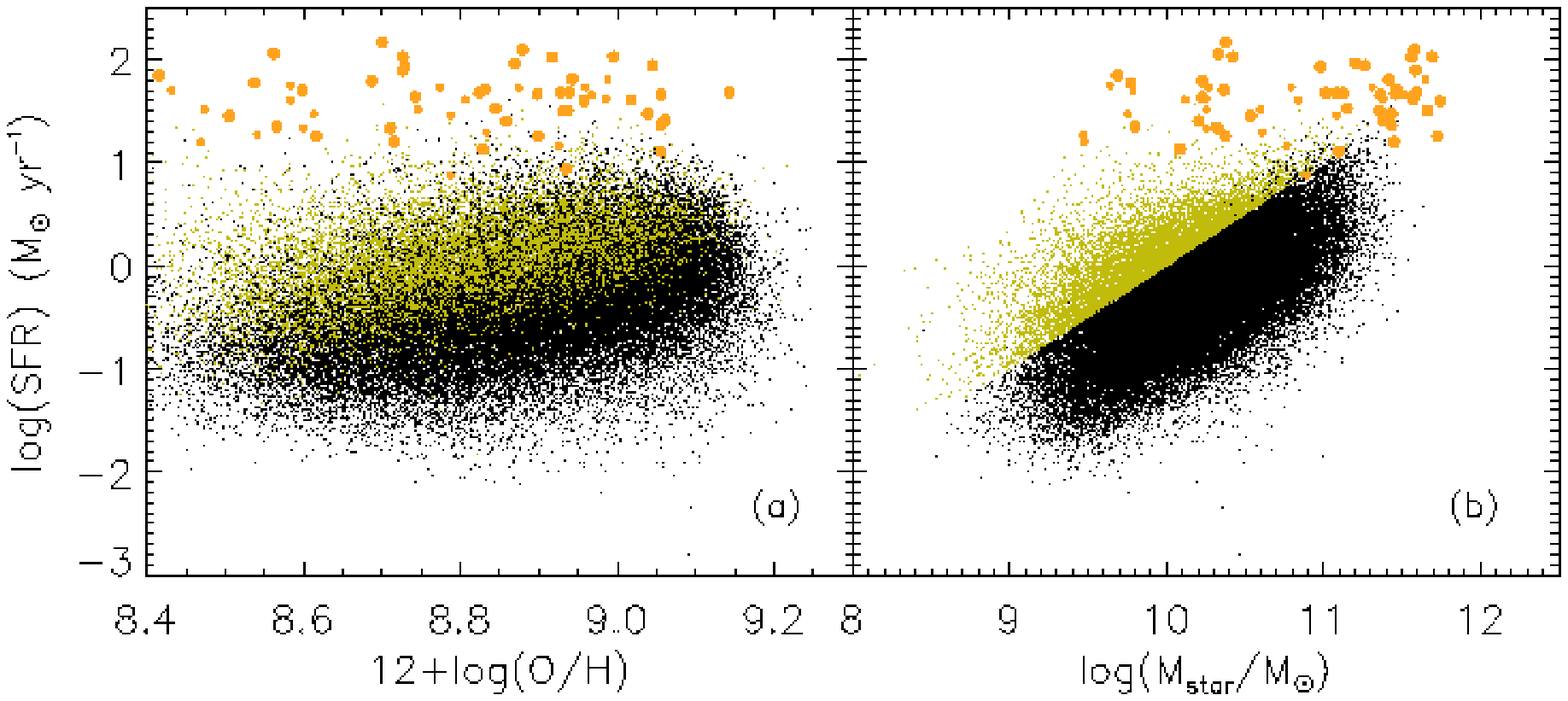}
\caption{Metallicity and Mass versus log(SSFR). Black dots and yellow circles represent galaxies at $z_0$ and $z_3$, respectively, while green dots, represent late-type galaxies at $z_0$ selected with log(SSFR) $> -10$.}
\end{figure*}

In this section, we investigate the morphology of the galaxies in our sample with the aim of clarify if  the $tail$ observed in Figs. 11 and 12, with higher SFR and SSFR in the $z_0$ sample, respectively, is related to specific morphological types. We will focus on samples $z_0$ and $z_3$ because those samples  show their galaxies uniformly distributed on mass. Galaxies of samples $z_1$ and $z_2$ show systematic problems due to sky lines, and to incompleteness, respectively.

We used the $g-r$ color, the concentration index c=R$_{90}$/R$_{50}$ (e.g. Park $\&$Choi et al. 2005), and the SSFR (e.g. Salim et al. 2009), which are the most common indexes to segregate early from late--type galaxies (see Fig. 13). We used k-corrected fiber colors for all the galaxies samples.

The color index has been commonly used as an early and late type morphological classificator (Baldry et al. 2004, Faber et al. 2007, Wang et al. 2007, Lee et al. 2007). Strateva et al. (2001) found that the integrated observed frame $u-r$ shows a bimodal distribution, however, they have shown that, when divided at $u-r=2.22$, the early and late type subsets have significant contamination, reaching about 30$\%$ for a sample with visually identified morphological types. Because the $u$ band shows large errors for the SDSS galaxies, we decided to use $g-r \lesssim 0.6$ (e.g. Schawinski et al. 2009, Masters et al. 2009), which allows separate early from late--type galaxies. As observed in Fig. 13a, $\sim$97$\%$ of the $z_3$ sample of galaxies correspond to a late--type morphology.

The concentration index c=R$_{90}$/R$_{50}$ has been successfully used in segregating late (c $<$ 2.86) from early--type (c $\geq$ 2.86)  subsets (e.g. Shimasaku et al. 2001, Strateva et al. 2001, Goto et al. 2002, Nakamura et al. 2003, Deng et al. 2009). Nevertheless, contamination in the early and late--type subsets separated using the concentration index, is typically about 20$\%$ (Yamauchi et al. 2005, Shimasaku et al. 2001). Using both, color and concentration index, Park $\&$ Choi et al. (2005) used the color-color space $u-r$ versus $\Delta(g-r)$ and the concentration index c$^{-1}$=R$_{50}$/R$_{90}$ $\sim$ 0.35 as a reliable morphological classificator.  In Fig. 13b, we show the concentration index c vs. $g-r$ for galaxies at $z_0$ and $z_3$, with the  $\sim$ 95$\%$ of the $z_3$ galaxies corresponding to late--type (c $<$ 2.86) galaxies.

Finally, the SSFR has been used as an indicator of early and late--type morphology (e.g. Wolf et al. 2009, Salim et al. 2009) since late--types have blue colors and  high SSFRs, while early--types have red colors and low SSFRs. For our $z_3$ sample, $\sim$ 89$\%$ of the galaxies have log(SSFR) $> -$10 (see Fig. 13c), and as reported by Salim et al. (2009), blue actively star-forming galaxies has log(SSFR) $> -$10, while lower values would correspond to the \emph{green valley} and red-sequence galaxies. As argued by Weinmann et al. (2006), the use of the SSFR would give us important clues in determining the morphological galaxy type, since for example, a genuine SF disk galaxy may appear red due to strong extinction (e.g. when seen edge-on), and thus be classified as early--type based on its color, while the SFR and morphology quantifiers would classify it as a late--type galaxy.

\begin{figure}[h]
\centering
\includegraphics[scale=0.7]{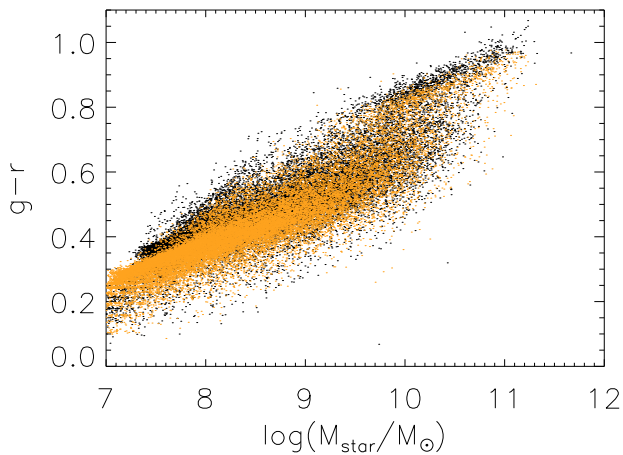}
\caption{Mass versus $g-r$ color for Millennium galaxies. Black and yellow dots represent galaxies at $z_0$ and $z_3$, respectively.}
\end{figure}

Our sample of galaxies at $z_3$ is mainly composed by late--type galaxies, as indicated by any of the morphological classificators discussed above. We decided to take these galaxies as a reference to delimitate the late--type zone. After trying all the discussed methods, in order to obtain a reliable morphological classification, and following the location of the $z_3$ galaxies in Fig. 13c, we conclude that the best method is to use both, log(SSFR) $> -10$, and a color $g-r$ $< 0.6$. After applying this criterion to our $z_0$ sample, we end with 7967 galaxies classified as late--type, corresponding to the $\sim$13$\%$ of the original sample. If we repeat the comparative analysis between SFR and metallicity, and SFR against stellar mass (see Fig. 14) for galaxies at $z_0$ and $z_1$, the $tail$ with higher SFR identified in previous sections corresponds to late--type galaxies. Note that the separation between late and early--type in the stellar mass versus SFR is a straight frontier (see Fig. 14b), because the separation criterion is the SSFR.

In order to alternatively assess this statement, we compare our results with selected mock galaxy samples from the Millennium simulations (Springel et al. 2005). We used the Bower2006a catalog (Bower et al. 2006), which give us redshift, SDSS k-corrected colors, stellar mass, and {H$\alpha$} luminosity, among other information. From the original catalog, we selected randomly 35000 galaxies at z=0 and z=0.4 with the {H$\alpha$} line in emission in order to be able to compare with our galaxies. As shown in Fig. 15, there is a clean  separation between a red sequence, formed by galaxies with $g-r$ $\gtrsim$ 0.8, the \emph{green valley} with 0.6 $\lesssim$ $g-r$ $\lesssim$ 0.8, and the blue cloud with $g-r$ $\lesssim$ 0.6. It can be observed how galaxies move towards late--type galaxies as redshift increases. Then, this results are consistent with observing mainly late--type galaxies at redshift $z_3$.

\section{Summary and Conclusions}

We analyzed a sample of emission line galaxies selected in four redshift intervals from $\sim$0 to 0.4 in bins of 0.1, taking into account the magnitude completeness of every redshift interval. In this paper we introduced the S2N2 diagram as a star-forming, composite, and AGNs galaxy classificator, we estimated metallicities using the R$_{23}$ method and analyzed the evolutive effects of galaxies from the three BPT diagrams. Additionally, we studied the evolution of the $M-Z$ and $L-Z$ relations, and analyzed the evolution and  implications of the galaxy morphology in the SFR--mass and metallicity relations. From these analysis we conclude the following:

\begin{itemize}
\item  Using the Kew01 photoionization grids, and the Kauf03 and Kew01 SF, and starburst limit  respectively, in the [{N\,\textsc{ii}}] /{H$\alpha$} vs [{O\,\textsc{iii}}] $\lambda$5007/{H$\beta$} diagram, we have demonstrated that the S2N2 is a well--behaved diagnostic diagram efficiently classifying star-forming, composite, and AGNs galaxies.

\item We  analyzed the galaxy evolution using the three main BPT diagrams: [{N\,\textsc{ii}}] /{H$\alpha$},  [{S\,\textsc{ii}}] /{H$\alpha$}, and [{O\,\textsc{i}}] $\lambda$6300/{H$\alpha$} vs  [{O\,\textsc{iii}}] $\lambda$5007/{H$\beta$} in our four redshift bins, observing an evolution toward higher values of the  [{O\,\textsc{iii}}] $\lambda$5007/{H$\beta$} ratio. This evolution is a consequence of the metallicity evolution as redshift increases, reflected in the three BPT diagrams, because the ratio  [{O\,\textsc{iii}}] $\lambda$5007/{H$\beta$}  is a good metallicity indicator. As a result, a metallicity decrement will be reflected in higher values of this ratio.

\item We analyzed the evolution of the $M-Z$ and $L-Z$ relations, observing that at higher redshift values, both relations evolve towards lower values of metallicity. We discovered that the flat zone of the $M-Z$ relation reported by Tremonti et al. (2004) for galaxies with log(M$_{star}$/M$_{\odot}$) $\gtrsim$ 10.5, is mainly constituted by  galaxies at $z > 0.1$ (samples at $z_1$, $z_2$ and $z_3$). Galaxies at $z_0$ redshift could be fitted with a linear function. Our $M-Z$ relation at redshift $z_3$ is $\sim$0.2 dex lower than our local one.

\item Our fit to the $M-Z$ relation for sample $z_3$ is in agreement with the one of Erb et al, (2006a) at $z\sim2.2$. We attribute this similarity to the galaxy morphology in the different samples, since our $z_3$ sample is conformed by late--type galaxies, while the sample of Erb et al. is composed by a mix of early and late--type galaxies. According to Calura et al. (2009), the $M-Z$ relation of late--type galaxies will have systematically lower metallicities than a $M-Z$ relation conformed by a mix of early and late--type galaxies.

\item  We analyzed the evolution of the mass-to-light ratio, observing lower $M/L$ ratios as redshift increase. For a small range of absolute magnitudes in the $z_3$ sample, we have a wide range of mass, making it possible to generate the $M-Z$ relation, but difficult to generate the $L-Z$ relation.

\item The decrement in metallicity observed in previous papers for galaxies at redshift $z_3$ (Lara-L\'opez et al. 2009a,b) is also observed, even though in this study we are not restricting our galaxy luminosities as in our previous studies.

\item We estimated the SFR and SSFR for our sample of galaxies and analyzed its relation with 12+log(O/H) and log(M$_{star}$/M$_{\odot}$), confirming the existence of a main sequence reported by Noeske et al. (2007) in the log(SFR) vs. log(M$_{star}$/M$_{\odot}$) plot. Consistently, we found that higher SFRs and SSFRs increase with redshift.

\item We analyzed the morphology of our galaxies through the $g-r$ color, the concentration index R$_{90}$/R$_{50}$, and the SSFR, concluding that the best method to determine the morphology was combining both, a color of $g-r < 0.6$, and a log(SSFR)$>$10 for selecting late--type galaxies.

\item Our $z_3$ sample of galaxies is mainly formed by late--type galaxies, a fact that helped us to classify morphological types at lower redshift. The fact that at higher redshift the fraction of late--type galaxies is larger, was confirmed by using mock galaxy catalogues from Millennium simulations.

\item We found at the higher redshift, a population with higher SFR and SSFR than the galaxies in the $z_0$ sample. After classifying late and early--type galaxies in the $z_0$ sample, we realized that the observed $tail$ showing higher SFR and SSFR is formed by late--type galaxies, demonstrating the connection of the galaxy morphology with the SFR in a new fashion.

\end{itemize}

Our work provide a useful tool for classifying galaxies with the S2N2 diagram, and demonstrating how galaxies evolve on the BPT diagrams as a consequence of metallicity evolution. We also analyzed the mass, metallicity and SFR relations, noting that galaxies in the redshift sample $z_3$ have lower values of metallicity, higher SFRs, and morphology indicators associated to late--types. In this study we pointed out the importance of the morphology of galaxies when deriving conclusions. Since a sample conformed by late--type galaxies will show lower values of metallicity than ones formed by a mix of morphological types.

\begin{acknowledgements}
This work was supported by the Spanish
\emph{Plan Nacional de Astronom\'{\i}a y Astrof\'{\i}sica} under grant AYA2008-06311-C02-01. The Sloan Digital Sky Survey (SDSS) is a joint project of The University of Chicago, Fermilab, the Institute for Advanced Study, the Japan Participation Group, The Johns Hopkins University, the Max--Planck--Institute for Astronomy, Princeton University, the United States Naval Observatory, and the University of Washington. Apache Point Observatory, site of the SDSS, is operated by the Astrophysical Research Consortium. Funding for the project has been provided by the Alfred P. Sloan Foundation, the SDSS member institutions, the National Aeronautics and Space Administration, the National Science Foundation, the U.S. Department of Energy, and Monbusho. The official SDSS web site is www.sdss.org. The Millennium Simulation databases used in this paper and the web application providing online access to them were constructed as part of the activities of the German Astrophysical Virtual Observatory. We thank the Starlight Project Team (UFSC, Brazil), specially to William Schoenell, who helped us downloading the whole data set. We thank to Romano Corradi for given us the idea to use the S2N2 diagram in galaxies. We thank to Kerttu Viironen for providing us the lines and metallicity data for several {H\,\textsc{ii}} and PNe to test the efficiency of the S2N2 diagram. Maritza A. Lara-L\'opez is supported by a CONACyT and SEP mexican fellowships.

\end{acknowledgements}


\begin{thebibliography}{}

   \bibitem[2003]{Abraham} Abraham, R. G., van den Bergh, S., Nair, P. 2003, ApJ, 588, 218

   \bibitem[2007]{Adelman--McCarthy} Adelman--McCarthy, J. K. et al. 2007, ApJs, 172, 634

   \bibitem[1942]{Aller}  Aller, L. H. 1942, ApJ, 95, 52

\bibitem[1979]{Alloin} Alloin, D., Collin--Souffrin, S., Joly, M., $\&$ Vigroux, L. 1979, A$\&$A, 78, 200 

\bibitem[2007]{Asari} Asari, N. V., Cid Fernandes R., Stasi\'nska G., et al. 2007, MNRAS, 381, 263

   \bibitem[1981]{Baldwin} Baldwin J., Phillips M., Terlevich R., 1981, PASP, 93, 5 (BPT)

   \bibitem[2004]{Baldry} Baldry, I, K., Glazebrook, K., Brinkmann, J., et al. 2004, ApJ, 600, 681

   \bibitem[2005]{Bauer} Bauer, A. E., Drory, N., Hill, G. J., $\&$ Feulner, G. 2005, ApJ, 621, L89

   \bibitem[2005]{Bell} Bell, E. F., Papovich, C., Wolf, C., et al. 2005, ApJ, 625, 23

   \bibitem[2005]{Bicker} Bicker, J. $\&$  Fritze-v. Alvensleben, U. 2005, A$\&$A, 443, L19

   \bibitem[2003]{Blanton} Blanton, M. R., Brinkmann, J., Csabai, I., et al. 2003, AJ, 125, 2348

  \bibitem[2006]{Bower} Bower, R. G., Benson, A. J., Malbon, R., et al. 2006, MNRAS, 370, 645

   \bibitem[2006]{Bresolin} Bresolin, F. 2006, preprint (astro-ph/0608410)

   \bibitem[2004]{Bresolin} Bresolin, F., Garnett, D. R., $\&$ Kennicutt, R. C. 2004, ApJ, 615, 228

   \bibitem[2004]{Brinchmann} Brinchmann, J., Charlot, S., White, S. D. M., et al. 2004, MNRAS, 351, 1151

   \bibitem[2000]{Brinchmann} Brinchmann, J., $\&$ Ellis, R. S. 2000, ApJ, 536, L77

   \bibitem[1991]{Brodie} Brodie, J. P., $\&$ Huchra, J. P. 1991, ApJ, 379, 157

   \bibitem[2007]{Brooks} Brooks, A. M., Governato, F., Booth, C. M., et al. 2007, ApJ, 655, L17

   \bibitem[2003]{Bruzual} Bruzual, G., Charlot S. 2003, MNRAS, 344, 1000 

   \bibitem[2008]{Buat} Buat, V., Boissier, S., Burgarella, D., et al. 2008, A$\&$A, 483, 107    

   \bibitem[2009]{Calura} Calura, F., Pipino, A., Chiappini, C., Matteucci, F.,  Maiolino, R. 2009, A$\&$A, 504, 373

   \bibitem[1981]{Canto} Cant\'o, J. 1981, in Investigating the Universe, ed. Z. Kopal $\&$ F. D. Kahn (Dordrecht: Reidel), 95

   \bibitem[2006]{Caputi} Caputi, K. I., Dole, H., Lagache, G., et al. 2006, ApJ, 637, 727

   \bibitem[1989]{Cardelli} Cardelli, J. A., Clayton G.C., Mathis J.S. 1989, ApJ, 345, 245    
   
   \bibitem[2001]{Carollo} Carollo, C. M., $\&$ Lilly, S. J. 2001, ApJ, 548, L153    

   \bibitem[2003]{Chabier} Chabrier, G. 2003, PASP, 115, 763

   \bibitem[2002]{Charlot} Charlot, S., Kauffmann, G., Longhetti, M., et al. 2002, MNRAS, 330, 876

  \bibitem[2007]{Cid Fernandes} Cid Fernandes R., Asari N. V., Sodr\'e L., et al. 2007, MNRAS, 375, L16   
             
   \bibitem[2005]{Cid Fernandes} Cid Fernandes, R., Mateus A., Sodr\'e L., Stasi\'nska G., Gomes J.M. 2005, MNRAS, 358, 363 

   \bibitem[1976]{Cohen} Cohen, J. G. 1976, ApJ, 203, 587

   \bibitem[2002]{Condon} Condon, J. J., Cotton, W. D., Broderick, J. J. 2002, AJ, 124, 675

   \bibitem[2006]{Conselice} Conselice, C. J., 2006, MNRAS, 373, 1389

   \bibitem[1996]{Cowie} Cowie, L. L., Songaila, A., Hu, E. M., Cohen, J. G. 1996, AJ, 112, 839

   \bibitem[2004]{De Lucia} De Lucia, G., Kauffmann, G., $\&$ White, S. D. M. 2004, MNRAS, 349, 1101

   \bibitem[1986]{Dekel} Dekel, A., $\&$ Silk, J. 1986, ApJ, 303, 39

   \bibitem[2009]{Deng} Deng, X., He, J., Wu, P., Ding, Y. 2009, ApJ., 699, 948

   \bibitem[2002]{Denicolo} Denicol\'o, G., Terlevich, R., $\&$  Terlevich, E. 2002, MNRAS, 330, 69

   \bibitem[2004]{Dickinson} Dickinson, M., Stern, D., Giavalisco, M., et al. 2004, ApJ, 600, L99

   \bibitem[1984]{Dopita} Donas, J., Deharveng, J. M. 1984,  A$\&$A, 140, 325

   \bibitem[1986]{Dopita} Dopita, M. A., $\&$ Evans, I. N. 1986, ApJ, 307, 431     

   \bibitem[2006]{Dopita} Dopita, M. A., Fischera, J., Sutherland, R. S., et al. 2006, ApJS, 167, 177

   \bibitem[2000]{Dopita} Dopita, M. A., Kewley, L. J., Heisler, C. A., $\&$ Sutherland, R. S. 2000, ApJ, 542, 224

   \bibitem[2002]{Dopita} Dopita, M. A., Periera, L., Kewley, L. J., $\&$ Capacciolo, M. 2002, ApJS, 143, 47

   \bibitem[1984]{Edmunds}  Edmunds, M. G., $\&$ Pagel, B. E. J. 1984, MNRAS, 211, 507

   \bibitem[2000]{Efstathiou}  Efstathiou, G. 2000, MNRAS, 317, 697

   \bibitem[2008]{Ellison} Ellison, S. L., Patton, D. R., Simard, L., McConnachie, A. W., 2008, ApJ, 672, L107

 \bibitem[2006]{Erb}  Erb, D. K., Shapley, A. E., Pettini, M., et al. 2006a, ApJ, 644, 813

 \bibitem[2006]{Erb}  Erb, D. K., Steidel, C. C., Shapley, A. E., et al. 2006b, ApJ, 646, 107

 \bibitem[2007]{Faber} Faber, S. M., Willmer, C. N. A., Wolf, C., et al. 2007, ApJ, 665, 265

 \bibitem[2005]{Feulner} Feulner, G., Gabasch, A., Salvato, M., et al. 2005, ApJ, 633, L9

 \bibitem[2008]{Finlator} Finlator, K., $\&$ Dav\'e, R. 2008, MNRAS, 385, 2181

 \bibitem[1977]{Fioc} Fioc, M., $\&$ Rocca-Volmerange, B. 1997, A$\&$A, 329, 950

   \bibitem[1989]{Gallagher} Gallagher, J. S., Hunter, D. A., $\&$ Bushouse, H. 1989, AJ, 97, 700

   \bibitem[1991]{Garcia-Lario} Garc\'{\i}a-Lario, O., Manchado, A., Riera, A., Mampaso, A., Pottasch, S. R. 1991, A$\&$A, 249, 223

   \bibitem[2004]{Garnett} Garnett, D. R., Kennicutt, R. C., $\&$ Bresolin, F. 2004, ApJ, 607, L21

   \bibitem[1987]{Garnett} Garnett, D. R., Shields, G. A. 1987, ApJ, 317, 82

   \bibitem[1997]{Garnett} Garnett, D. R., Shields, G. A., Skillman, E. D., Sagan, S. P., $\&$ Dufour, R. J. 1997, ApJ, 489, 36

   \bibitem[2002]{Gavazzi} Gavazzi, G., Bonfanti, C., Sanvito, G., Boselli, A., $\&$ Scodeggio, M. 2002, ApJ, 576, 135

   \bibitem[1996]{Gavazzi} Gavazzi, G., Scodeggio, M., 1996, A$\&$A, 312, L29

   \bibitem[2002]{Goto} Goto, T., Okamura, S., McKay, T. A., et al. 2002, PASJ, 54, 515

   \bibitem[2006]{Gunn} Gunn, J. E., Siegmund, W. A., Mannery, E. J., et al. 2006, AJ, 131, 2332

   \bibitem[1997]{Guzman} Guzm\'an, R., Gallego, J., Koo, D. C., et al. 1997, ApJ, 489, 559

   \bibitem[1973]{Harper} Harper, D. A., $\&$ Low, F. J. 1973, ApJ, 182, L89

   \bibitem[2005]{Hammer} Hammer, F., Flores, H., Elbaz, D., et al. 2005,  A$\&$A, 430, 115

   \bibitem[2000]{Henry} Henry, R. B. C., Edmunds, M. G., $\&$ K\"oppen, J. 2000, ApJ, 541, 660

   \bibitem[2007]{Henry} Henry, R. B. C., Prochaska Jason X. 2007, PASP, 119, 962 

   \bibitem[1926]{Hubble} Hubble E. 1926, ApJ, 64, 321

   \bibitem[1986]{Hunter} Hunter, D. A. $\&$ Gallagher, J. S., III. 1986, PASP, 98, 5

   \bibitem[2000]{Jansen} Jansen, R. A., Fabricant, D., Franx, M., $\&$ Caldwell, N. 2000, ApJS, 126, 331

   \bibitem[2001]{Jansen} Jansen, R. A., Franx, M., $\&$ Fabricant, D. 2001, ApJ, 551, 825

   \bibitem[2005]{Juneau} Juneau, S., Glazebrook, K., Crampton, D., et al. 2005, ApJ, 619, L135

   \bibitem[2003]{Kauffmann} Kauffmann, G., Heckman, T. M., Tremonti, C., et al. 2003a, MNRAS, 346, 1055 (Kauf03)

   \bibitem[2003]{Kauffmann} Kauffmann, G.,Heckman, T. M., White, S. D., et al. 2003b, MNRAS, 341, 54

   \bibitem[2004]{Kauffmann} Kauffmann, G., White, S. D. M., Heckman, T. M., et al. 2004, MNRAS, 353, 713

   \bibitem[1998]{Kennicutt} Kennicutt, R. C. 1998, ARA$\&$A, 36, 189 

   \bibitem[1994]{Kennicutt} Kennicutt, R. C., Jr., Tamblin, P. $\&$ Congdon, C. 1994, 435, 22

   \bibitem[2003]{Kennicutt} Kennicutt, R. C., Jr., Bresolin, F., $\&$ Garnett, D. R. 2003, ApJ, 591, 801     

   \bibitem[1983]{Kennicutt} Kennicutt, R. C., $\&$ Kent, S. M. 1983, AJ, 88, 1094 

   \bibitem[2007]{Kewley} Kewley, L. J., Brown, W. R., Geller, M. J., Kenyon, S. J., $\&$ Kurtz, M. J. 2007, AJ, 133, 882

   \bibitem[2001]{Kewley} Kewley, L. J., Dopita M. A., Sutherland R. S., Heisler C. A., Trevena J. 2001, ApJ, 556, 121 (Kew01)
      
   \bibitem[2002]{Kewley} Kewley, L. J., $\&$ Dopita, M. A. 2002, ApJS, 142, 35   

   \bibitem[2008]{Kewley} Kewley, L. J., $\&$ Ellison, S. L. 2008, ApJ, 681, 1183   

   \bibitem[2004]{Kewley} Kewley, L. J., Geller, M. J., Jansen, R. A., 2004, AJ, 127, 2002

   \bibitem[2006]{Kewley} Kewley, L. J., Groves, B., Kauffmann, G., $\&$ Heckman, T. 2006, MNRAS, 372, 961
   
   \bibitem[2005]{Kewley} Kewley, L. J., Jansen, R. A., $\&$ Geller, M. J. 2005, PASP, 117, 227   

   \bibitem[1981]{Kinman} Kinman, T. D., $\&$ Davidson, K. 1981, ApJ, 243, 127

   \bibitem[2007]{Kobayashi} Kobayashi, C., Springel, V., $\&$ White, S. D. M. 2007, MNRAS, 376, 1465

   \bibitem[2004]{Kobulnicky} Kobulnicky, H. A., $\&$ Kewley, L. J. 2004, ApJ, 617, 204

   \bibitem[2003]{Kobulnicky} Kobulnicky, H. A., Willmer, C. N. A., Phillips, A. C., et al. 2003, ApJ, 599, 1006

   \bibitem[2007]{Koppen} K\"oppen, J., Weidner, C., Kroupa, P. 2007, MNRAS, 375, 673

   \bibitem[2001]{Kroupa} Kroupa, P. 2001, MNRAS, 322, 231

   \bibitem[2009]{Lamareille} Lamareille, F., Brinchmann, J., Contini, T., et al. 2009, A$\&$A, 495, 53 

   \bibitem[2006]{Lamareille} Lamareille, F., Contini, T., Brinchmann, J., et al. 2006, A$\&$A, 448, 907    

   \bibitem[2004]{Lamareille} Lamareille, F., Mouhcine, M., Contini, T., Lewis, I., $\&$ Maddox, S. 2004, MNRAS, 350, 396

   \bibitem[2009]{Lara-Lopez} Lara-L\'opez, M. A., Cepa, J., Bongiovanni, A., et al. 2009a, A$\&$A, 493, L5

   \bibitem[2009]{Lara-Lopez} Lara-L\'opez, M. A., Cepa, J., Bongiovanni, A., et al. 2009b, A$\&$A, 505, 529

   \bibitem[1974]{Larson} Larson, R. B. 1974, MNRAS, 169, 229

   \bibitem[1978]{Larson} Larson, R. B., $\&$ Tinsley, B. M. 1978, ApJ, 219, 46

   \bibitem[2006]{Lee} Lee J. C. 2006, PhD. thesis, Univ. Arizona

   \bibitem[2007]{Lee} Lee J. H., Lee M. G., Kim T., et al, 2007, ApJ, 663, L69

   \bibitem[1999]{Leitherer} Leitherer, C., Schaerer, D., Goldader, J. D., et al. 1999, ApJS, 123, 3

   \bibitem[1979]{Lequeux} Lequeux, J., Peimbert, M., Rayo, J. F., Serrano, A.,  $\&$ Torres-Peimbert, S. 1979, A$\&$A, 80, 155

   \bibitem[2010]{Levesque} Levesque, E. M., Kewley, L. J., Larson, K. L. 2010, ApJ, 139, 712

   \bibitem[2006]{Liang} Liang, Y. C., Yin, S. Y., Hammer, F., et al. 2006, ApJ, 652, 257

   \bibitem[2007]{Liang} Liang, Y. C., Hammer, F., Yin, S. Y., et al. 2007, A$\&$A, 473, 411

   \bibitem[2003]{Lilly} Lilly , S. J., Carollo, C. M.,  $\&$ Stockton, A. 2003, ApJ, 597, 730

   \bibitem[2008]{Lintott} Lintott, C. J., Schawinski, K., Slosar, A., et al. 2008, MNRAS, 389, 1179

   \bibitem[2008]{Liu} Liu, X., Shapley, A. E., Coil, A. L., Brinchmann, J., $\&$ Ma, C.-P. 2008, ApJ, 678, 758

   \bibitem[1999]{MacLow}  MacLow, M., $\&$ Ferrara, A. 1999, ApJ, 513, 142

   \bibitem[2003]{Magrini}  Magrini, L., Perinotto, M., Corradi, R. L. M., Mampaso, A. 2003, A$\&$A, 400, 511

   \bibitem[2008]{Maiolino}  Maiolino, R., Nagao, T., Grazian, A., et al. 2008, A$\&$A, 488, 463

   \bibitem[2004]{Maier} Maier, C., Meisenheimer, K., $\&$ Hippelein, H. 2004,  A$\&$A, 418, 475

   \bibitem[2006]{Maier} Maier, C., Lilly, S., Carollo, C. M., et al. 2006, ApJ, 639, 858     
   
   \bibitem[2005]{Maier} Maier, C., Lilly, S., Carollo, C. M., Stockton, A., $\&$ Brodwin, M. 2005, ApJ, 634, 849       

   \bibitem[2009]{Mannucci} Mannucci, F., Cresci, G., Maionilo, R., et al. 2009, MNRAS, 398, 1915

   \bibitem[2009]{Masters} Masters, K. L., Mosleh, M., Romer, A. K., et al. 2009, MNRAS, preprint (arXiv:0910.4113))

   \bibitem[2006]{Mateus} Mateus, A., Sodr\'e L., Cid Fernandes R., et al. 2006, MNRAS, 370,721 
   
   \bibitem[1985]{McCall} McCall, M. L., Rybski, P. M., $\&$ Shields, G. A. 1985, ApJS, 57, 1      

  \bibitem[1968]{McClure} McClure, R. D., van den Bergh, S., 1968, AJ, 73, 1008

  \bibitem[1991]{McGaugh} McGaugh, S. S. 1991, ApJ, 380, 140

  \bibitem[2002]{Melbourne}  Melbourne, J., $\&$ Salzer, J.J. 2002, AJ, 123, 2302

  \bibitem[2009]{Mobasher} Mobasher, B., Dahlen, T., Hopkins, A., et al. 2009, ApJ, 690, 1074

  \bibitem[2006]{Moustakas} Moustakas, J., Kennicutt, R. C. Jr., 2006, ApJ, 651, 155

  \bibitem[2008]{Mouhcine} Mouhcine, M., Gibson, B. K., Renda, A., $\&$ Kawata, D. 2008, A$\&$A, 486, 711

   \bibitem[2006]{Nagao} Nagao, T., Maiolino, R., $\&$ Marconi, A. 2006, A$\&$A, 459, 85

   \bibitem[2003]{Nakamura} Nakamura, O., Fukugita, M., Yasuda, N., et al. 2003, AJ, 125, 1682

   \bibitem[2007]{Noeske} Noeske, K. G., Weiner, B. J., Faber, S. M., et al. 2007a, ApJ, 660, L43

   \bibitem[2007]{Noeske} Noeske, K. G., Faber, S. M., Weiner, B. J., et al. 2007b, ApJ, 660, L47

   \bibitem[1989]{Osterbrock} Osterbrock D.R., 1989, Astrophysics of Gaseous Nebulae and Active Galactic Nuclei. University Science Books, Mill Valley CA     
      
   \bibitem[1979]{Pagel} Pagel, B. E. J., Edmunds, M. G., Blackwell, D. E., Chun, M. S., $\&$ Smith, G. 1979, MNRAS, 189, 95   

   \bibitem[1986]{Pagel} Pagel, B. E. J., 1986, PASP, 98, 1009 
   
   \bibitem[1992]{Pagel} Pagel, B. E. J., Simonson, E. A., Terlevich, R. J., $\&$ Edmunds, M. G. 1992, MNRAS, 255, 325

   \bibitem[2009]{Pannella} Pannella, M., Carilli, C. L., Daddi, E., et al. 2009, ApJ, 698, L116

   \bibitem[2006]{Papovich} Papovich, C., Moustakas, L. A., Dickinson, M., et. al. 2006, ApJ, 640, 92

   \bibitem[2005]{Park} Park, C.,  $\&$ Choi, Y. 2005, ApJ, 635, L29

   \bibitem[2009]{Perez-Montero} P\'erez-Montero, E., $\&$ Contini, T. 2009, MNRAS, 398, 949

   \bibitem[2005]{Perez-Gonzalez} P\'erez-Gonz\'alez, P. G., Rieke, G. H., Egami, E., et al. 2005, ApJ, 630, 82

   \bibitem[2002]{Pettini} Pettini, M., Ellison, S. L., Bergeron, J.,  $\&$ Petitjean, P. 2002,  A$\&$A, 391, 21

    \bibitem[2004]{Pettini} Pettini, M.,  $\&$ Pagel, B. E. J. 2004,  MNRAS, 348, L59
   
   \bibitem[2000]{Pilyugin} Pilyugin, L. S. 2000, A$\&$A, 362, 325   

   \bibitem[2001]{Pilyugin} Pilyugin, L. S. 2001, A$\&$A, 369, 594  
	
   \bibitem[2000]{Pilyugin} Pilyugin, L. S., Ferrini, F., 2000, A$\&$A, 358, 72

   \bibitem[2005]{Pilyugin} Pilyugin, L. S., $\&$ Thuan, T. X. 2005, ApJ, 631, 231

    \bibitem[2000]{Raimann} Raimann, D., Storchi-Bergmann, T., Bica, E., Melnick, J., $\&$ Schmitt, H. 2000, MNRAS, 316, 559

    \bibitem[2006]{Reddy} Reddy, N. A., Steidel, C. C., Fadda, D., et al. 2006, ApJ, 644, 792

    \bibitem[1995]{Richer} Richer, M. G., McCall, M. L., 1995, ApJ, 445, 642

    \bibitem[1978]{Rieke} Rieke, G. H., Lebofsky, M. J., 1978, ApJ, 220, L37

    \bibitem[2005]{Riesgo} Riesgo, H., $\&$ L\'opez, J. A. 2005, Rev. Mex. Astron. Astrofis., 41, 57

    \bibitem[2006]{Riesgo} Riesgo, H., $\&$ L\'opez, J. A. 2006, Rev. Mex. Astron. Astrofis., 42, 47

    \bibitem[1963]{Roberts} Roberts, M. S. 1963, ARAA, 1, 149

    \bibitem[2008]{Rodrigues} Rodrigues, M., Hammer, F., Flores, H., et al. 2008, A$\&$A, 492, 371

    \bibitem[2002]{Rosa-Gonzalez} Rosa-Gonz\'alez, D., Terlevich, E., $\&$ Terlevich, R. 2002, MNRAS, 332, 283

    \bibitem[2009]{Rosa-Gonzalez} Rosa Gonz\'alez, D., Terlevich, E., Jim\'enez Bail\'on, E., et al. 2009, MNRAS, 399, 487

    \bibitem[2009]{Rovilos} Rovilos, E., Georgantopoulos, I., Tzanavaris, P., et al. 2009, A$\&$A, 502, 85

    \bibitem[2005]{Savaglio} Savaglio, S., Glazebrook, K., Le Borgne, D., et al. 2005, ApJ, 635, 260

    \bibitem[1977]{Sabbadin} Sabbadin, F., Minello, S., $\&$ Bianchini, A. 1977, A$\&$A, 60, 174

    \bibitem[2009]{Salim} Salim, S., Dickinson, M., Rich, R. M., et al. 2009, ApJ, 700, 161

    \bibitem[2005]{Salim} Salim, S., Charlot, S., Rich, M., et al. 2005, ApJ, 619, L39

    \bibitem[1961]{Sandage} Sandage A. R., 1961, The Hubble Atlas of Galaxies. Carnegie Institute of Washington, Washington

    \bibitem[2005]{Savaglio} Savaglio, S., Glazebrook, K., Le Borgne., et al. 2005, ApJ, 635, 260 

   \bibitem[2008]{Scannapieco} Scannapieco, C., Tissera, P. B., White, S. D. M., Springel, V., 2008, MNRAS, 389, 1137 

    \bibitem[2007]{Scarlata} Scarlata, C., Carollo, C. M., Lilly, S., et al. 2007, ApJS, 172, 406

   \bibitem[2009]{Schawinski} Schawinski, K., Virani, S., Simmons, B., et al. 2009, ApJ, 692, L19

   \bibitem[1998]{Schlegel} Schlegel, D.J., Finkbeiner D.P., Davis M. 1998, ApJ, 500, 525

   \bibitem[1971]{Searle} Searle, L. 1971, ApJ, 168, 327 

   \bibitem[1973]{Searle} Searle, L., Sargent, W. L. W., $\&$ Bagnuolo, W. G. 1973, ApJ, 179, 427

   \bibitem[2002]{Serjeant} Serjeant S., Gruppioni C., Oliver S., 2002, MNRAS, 330, 621

   \bibitem[2005]{Shapley} Shapley, A. E., Steidel, C. C., Erb, D, K., et al. 2005, ApJ, 626, 698

   \bibitem[1990]{Shields} Shields, G. A. 1990, ARA$\&$A, 28, 525 

   \bibitem[2001]{Shimasaku} Shimasaku, K., Fukugita, M., Doi, M., al. 2001, AJ, 122, 1238

   \bibitem[1993]{Skillman} Skillman, E. D., $\&$ Kennicutt, R. C., Jr. 1993, ApJ, 411, 655

   \bibitem[1989]{Skillman} Skillman, E. D., $\&$ Kennicutt, R. C., Jr., $\&$ Hodge, P. W.  1989, ApJ, 347, 875

   \bibitem[2005]{Springel} Springel, V.,White, S. D. M., Jenkins, A., et al. 2005, Nature, 435, 629

   \bibitem[2006]{Stasinska} Stasi\'nska, G., Cid Fernandes, R., Mateus, A., Sodr\'e, L., $\&$ Asari, N. V. 2006, MNRAS, 371, 972
         
   \bibitem[2002]{Stoughton} Stoughton, C., Lupton, R. H., Bernardi, M., et al. 2002, AJ, 123, 485

   \bibitem[2002]{Strauss} Strauss, M. A., Weinberg, D. H., Lupton, R. H., et al. 2002, AJ, 124, 1810

   \bibitem[2001]{Strateva} Strateva, I., Ivezi\'c, Z., Knapp, G. R., et al. 2001, AJ, 122, 1861

   \bibitem[2008]{Tassis} Tassis, K., Kravtsov, A. V., $\&$ Gnedin, N. Y. 2008, ApJ, 672, 888

   \bibitem[1980]{Telesco} Telesco, C. M., Harper, D. A. 1980, ApJ, 235, 392

   \bibitem[1996]{Thurston} Thurston, T. R., Edmunds, M. G., $\&$ Henry, R. B. C. 1996, MNRAS, 283, 990

   \bibitem[1968]{Tinsley} Tinsley, B. M. 1968, ApJ, 151, 547

   \bibitem[1972]{Tinsley} Tinsley, B. M. 1972, A$\&$A, 20, 383

   \bibitem[2004]{Tremonti} Tremonti, C. A., Heckman, T. M., Kauffmann, G., et al. 2004, ApJ, 613, 898 (T04)

   \bibitem[1998]{Tresse} Tresse, L., Maddox, S. J. 1998, ApJ, 495, 691

   \bibitem[2009]{Vale Asari} Vale Asari N., Cid Fernandes R., Gomes J. M., et al.  2009, MNRAS, 396, L71

   \bibitem[1991]{de Vaucouleurs} de Vaucouleurs, G., de Vaucouleurs, A., Corwin, H. G., et al. 1991, Third Reference Catalog of Bright Galaxies., Springer-Verlag, New York

   \bibitem[1987]{Veilleux} Veilleux, S., $\&$ Osterbrock, D. E. 1987, ApJS, 63, 295 

    \bibitem[2007]{Viironen}  Viironen, K., Delgado-Inglada, G., Mampaso, A., Magrini, L., Corradi, R. L. M., 2007, MNRAS, 381, 1719

   \bibitem[1993]{Vila-Costas} Vila-Costas, M. B., $\&$ Edmunds, M. G. 1993, MNRAS, 256, 199
   	
   \bibitem[1996]{Vilchez} V\'{\i}lchez, J. M.,  $\&$  Esteban, C. 1996, MNRAS, 280, 720 

   \bibitem[2007]{Wang} Wang, Y., Yang, X., Mo, H. J., van den Bosch, F. C. 2007, ApJ, 664, 608

   \bibitem[2006]{Weinmann} Weinmann, A. M., van den Bosch, F. C., Yang, X., Mo, H. J. 2006, MNRAS, 366, 2

   \bibitem[2009]{Wolf} Wolf, C., Arag\'on-Salamanca, A., Balogh, M. 2009, MNRAS. 393, 1302

   \bibitem[2005]{Yamauchi} Yamauchi, C., $\&$ Goto, T. 2005, MNRAS, 359, 1557

   \bibitem[1994]{Zaritsky} Zaritsky, D., Kennicutt, R. C., $\&$ Huchra, J. P. 1994, ApJ, 420, 87 


 \end{thebibliography}
\end{document}